\documentclass[
reprint,
nofootinbib,
amsmath,amssymb,
aps,
prd,
floatfix,
showkeys
]{revtex4-2}

\usepackage[USenglish]{babel}

\usepackage{amsmath}
\usepackage{amssymb}
\usepackage{amsfonts}
\usepackage{amsbsy}
\usepackage{bm}            
\usepackage{mathrsfs}
\usepackage{slashed}
\usepackage{extarrows}

\usepackage{graphicx}
\usepackage{graphics}
\usepackage{dcolumn}       

\usepackage[dvipsnames]{xcolor}

\usepackage{array}
\usepackage{multirow}
\usepackage{enumitem}
\usepackage{booktabs}

\usepackage[caption=false]{subfig}
\usepackage{placeins}

\usepackage{url}
\usepackage[normalem]{ulem}         
\usepackage{verbatim}
\usepackage{latexsym}
\usepackage{orcidlink}
\usepackage{setspace}

\newcounter{widefn}

\PassOptionsToPackage{
	colorlinks,
	citecolor=blue,
	urlcolor=magenta,
	linkcolor=blue
}{hyperref}
\usepackage{hyperref}

\begin{document}

\title{\bf Testing Matter Diffusion with Late-Time Cosmological Observations}

\author{Alireza Zafari \orcidlink{0009-0004-8472-7609}}
\email{alirezazafari12@gmail.com}
\affiliation{Department of Physics, Faculty of Science,
Shahid Beheshti University, Tehran, Iran}

\author{Mohammad-Hossein Namdar\orcidlink{0009-0005-5289-8881}}
\email{mnamdar0@estudiante.uc.cl}
\affiliation{Instituto de Astrofísica, Facultad de Física,
Pontificia Universidad Católica de Chile,
Av. Vicuña Mackenna 4860, 7820436 Macul, Santiago, Chile}
\affiliation{Centro de Astro-Ingeniería UC,
Pontificia Universidad Católica de Chile,
Av. Vicuña Mackenna 4860, 7820436 Macul, Santiago, Chile}

\author{Luis A. Escamilla}
\email{torresl@itu.edu.tr}
\affiliation{Department of Physics, Istanbul Technical University, Maslak 34469 Istanbul, Turkey}

\author{Eleonora Di Valentino \orcidlink{0000-0001-8408-6961}}
\email{e.divalentino@sheffield.ac.uk}
\affiliation{School of Mathematical and Physical Sciences, University of Sheffield, Hounsfield Road,
Sheffield S3 7RH, United Kingdom}

\date{\today}

	
\begin{abstract}
We investigate a class of late-time cosmological models derived from the phenomenological framework of variable matter diffusion. In these scenarios, energy-momentum conservation requires a continuous energy exchange between matter and an effective scalar-field dark-energy component, $\phi$. We consider a baseline constant-diffusion scenario alongside four non-linear power-law parametrizations, in which the diffusion coefficient evolves as a function of the scale factor, matter density, scalar-field density, or Hubble expansion rate. To assess their cosmological viability, we implement these models within the \texttt{SimpleMC} inference framework and constrain them using late-time background observations, including cosmic chronometers (\texttt{CC}), \texttt{DESI\,DR2} baryon acoustic oscillation measurements, Type Ia supernovae from \texttt{PantheonPlus}, and the locally calibrated \texttt{PantheonPlus\&SH0ES} sample. In the absence of the local calibration ($\mathrm{CC}+\mathrm{DESI\,DR2}+\mathrm{PantheonPlus}$), the diffusion models yield only modest improvements in the fit ($\Delta\chi^2 \approx -4$), performing comparably to the two-parameter CPL parametrization while exhibiting negative Bayesian log-evidence differences relative to $\Lambda$CDM. In contrast, including the SH0ES calibration ($\mathrm{CC}+\mathrm{DESI\,DR2}+\mathrm{PantheonPlus\&SH0ES}$) leads to a dramatic reduction in the minimum $\chi^2$ ($\Delta\chi^2 \approx -22$) and decisive Bayesian evidence in favor of the diffusion framework ($\Delta\ln\mathcal{Z} \approx 9$). Remarkably, the single-parameter constant-diffusion model accounts for virtually all of this statistical improvement, outperforming the CPL parametrization by more than 10 units in $\chi^2$ and more than 8 units in $\Delta\ln\mathcal{Z}$. The four non-linear extensions provide only negligible additional improvements ($\Delta\chi^2 \approx -1$), indicating that current late-time background observations favor the presence of a non-vanishing matter-diffusion interaction over $\Lambda$CDM, while providing no statistically significant evidence for a specific time-dependent functional form of the diffusion coefficient.
\end{abstract}
	
\maketitle

\section{Introduction}\label{sec:intro}
	
The standard model of cosmology, usually referred to as $\Lambda$CDM, has been remarkably successful in describing the observable Universe~\cite{AtacamaCosmologyTelescope:2025blo, Planck:2018vyg, SPT-3G:2025bzu, eBOSS:2020fvk, Brout:2022vxf}. Its success relies on a minimal set of parameters that simultaneously accounts for the observed late-time accelerated expansion, cosmic microwave background (CMB) anisotropies, and the large-scale structure of the Universe. Nevertheless, the underlying physical nature of the dark sector remains fundamentally unknown. In particular, while treating dark energy as a cosmological constant ($\Lambda$) ensures theoretical simplicity, it introduces profound fine-tuning challenges~\cite{Weinberg:2000yb, Zlatev:1998tr, Carroll:2000fy, Peebles:2002gy} and remains phenomenologically limited. This motivates the exploration of extensions to $\Lambda$CDM in which the dark sector possesses a richer dynamical structure.

Beyond these theoretical considerations, several observational results have also motivated renewed interest in extensions of the standard cosmological model. Most notably, the Hubble tension~\cite{Verde:2019ivm,DiValentino:2020zio,DiValentino:2021izs,Perivolaropoulos:2021jda,Schoneberg:2021qvd,Shah:2021onj,Abdalla:2022yfr,DiValentino:2022fjm,Kamionkowski:2022pkx,Giare:2023xoc,Hu:2023jqc,Verde:2023lmm,DiValentino:2024yew,CosmoVerseNetwork:2025alb,Ong:2025cwv,Cai:2026swf}, namely the discrepancy between the value of the Hubble constant inferred from early-Universe probes~\cite{Planck:2018vyg,Planck:2018nkj,ACT:2025fju,SPT-3G:2025bzu} within the $\Lambda$CDM model and that measured using the local distance ladder~\cite{Freedman:2020dne,Birrer:2020tax,Riess:2021jrx,Anderson:2023aga,Scolnic:2023mrv,Jones:2022mvo,Anand:2021sum,Freedman:2021ahq,Uddin:2023iob,Huang:2023frr,Li:2024yoe,Pesce:2020xfe,Kourkchi:2020iyz,Schombert:2020pxm,Blakeslee:2021rqi,deJaeger:2022lit,Murakami:2023xuy,Breuval:2024lsv,Freedman:2024eph,Riess:2024vfa,Vogl:2024bum,Scolnic:2024hbh,Said:2024pwm,Boubel:2024cqw,Scolnic:2024oth,Li:2025ife,Jensen:2025aai,Riess:2025chq,Benisty:2025tct,Newman:2025gwg,Stiskalek:2025ktq,Agrawal:2025tuv,Bhardwaj:2025kbw}, has continued to increase in significance. The latest H$_0$ Distance Network determination~\cite{H0DN:2025lyy} reports a discrepancy exceeding $7\sigma$, reinforcing the possibility that extensions beyond the minimal $\Lambda$CDM model may be required. In parallel, recent analyses of the DESI baryon acoustic oscillation measurements~\cite{DESI:2025fii,DESI:2024mwx,DESI:2025zgx,DESI:2024kob}, particularly when combined with Type Ia supernova data~\cite{Hoyt:2026fve,DES:2025sig,Scolnic:2021amr}, have provided intriguing indications that the dark energy equation of state may deviate from that of a cosmological constant. While the statistical significance depends on the adopted supernova calibration and the combination of cosmological probes, several independent analyses have reported a preference for dynamical dark energy over $\Lambda$CDM at the $3$--$4\sigma$ level~\cite{DESI:2025fii,DESI:2024mwx,DESI:2025zgx,DESI:2024kob,Cortes:2024lgw,Shlivko:2024llw,Luongo:2024fww,Gialamas:2024lyw,Wang:2024dka,Ye:2024ywg,Tada:2024znt,Carloni:2024zpl,Chan-GyungPark:2024mlx,Bhattacharya:2024hep,Reboucas:2024smm,Najafi:2024qzm,Giare:2024gpk,Giare:2024ocw,Jiang:2024xnu,RoyChoudhury:2024wri,Giare:2024oil,Giare:2025pzu,Kessler:2025kju,RoyChoudhury:2025dhe,Scherer:2025esj,Wolf:2025jlc,Santos:2025wiv,Specogna:2025guo,Cheng:2025lod,Cheng:2025hug,Ozulker:2025ehg,Li:2025vuh,Lee:2025pzo,Fazzari:2025lzd,Smith:2025icl,Herold:2025hkb,Cheng:2025yue,Gokcen:2026pkq,Ishak:2025cay,Najafi:2026kxs,Yang:2026yaq,Kessler:2026dbi,Lee:2026yzs,Li:2026asg,Giare:2026oti,GuptaChoudhury:2026gsl}. Although these results are not yet conclusive, they have stimulated considerable interest in physically motivated scenarios capable of modifying the late-time cosmic expansion history while remaining compatible with current cosmological observations.
	
In this spirit, a prominent class of extensions investigates the non-gravitational exchange of energy and momentum within the dark sector. Commonly known as Interacting Dark Energy (IDE) models, these frameworks have been widely studied in the literature~\cite{Amendola:1999er, Zimdahl:2001ar, Farrar:2003uw, Wang:2016lxa, Wang:2024vmw, DiValentino:2019ffd, Montani:2024pou, vanderWesthuizen:2025rip, Salvatelli:2013wra, Kumar:2016zpg, Caprini:2016qxs, Murgia:2016ccp, Zheng:2017asg, Kumar:2017dnp, DiValentino:2017iww, Kumar:2021eev, Gao:2021xnk, Pan:2023mie, Benisty:2024lmj, Yang:2020uga, Forconi:2023hsj, Pourtsidou:2016ico, DiValentino:2020vnx, DiValentino:2020leo, Nunes:2021zzi, Yang:2018uae, Zhang:2018mlj, vonMarttens:2019ixw, Lucca:2020zjb, Xiao:2021nmk, Gao:2022ahg, Zhai:2023yny, Joseph:2022khn, Bernui:2023byc, Becker:2020hzj, Hoerning:2023hks, Giare:2024ytc, Mukhopadhyay:2020bml, Escamilla:2023shf, vanderWesthuizen:2023hcl, Silva:2024ift, Zhao:2022ycr, Li:2024qso, Pooya:2024wsq, Halder:2024uao, Castello:2023zjr, Yao:2023jau, Mishra:2023ueo, Nunes:2016dlj, Silva:2025hxw, Yang:2025uyv, Zhang:2025dwu, Li:2025muv, Li:2025owk, Li:2026xaz, Zhai:2026uwr}. In the standard cosmological paradigm, each sector—namely matter and dark energy—is assumed to be separately conserved. However, general relativity requires only that the \emph{total} energy-momentum tensor be covariantly conserved. Phenomenological models exploit this geometric freedom by introducing a coupling term that governs the energy transfer between the individual species. Matter-diffusion cosmology represents one such alternative framework~\cite{Calogero:2011re, Calogero:2012kd, Calogero:2013zba, Velten:2014uva, Haba:2016swv, Escamilla:2023shf}. Within this framework, the matter component is not independently conserved, transferring energy to a coupled sector that acts as an effective dark energy component.
	
In these models, matter particles undergo a velocity diffusion process that directly modifies the standard matter continuity equation~\cite{dudley1966lorentz, Herrmann:2010jn, Calogero:2011re, Calogero:2012kd, Calogero:2013zba}. To guarantee the covariant conservation of the total energy-momentum tensor, a cosmological scalar field $\phi$ is introduced to compensate for the energy exchanged by the diffusing matter component. This scalar field is conventionally identified as the effective dark energy sector. Consequently, this scenario provides a physically motivated interacting dark sector model wherein the late-time background expansion history naturally deviates from the standard baseline, while the standard $\Lambda$CDM paradigm is seamlessly recovered in the limit of vanishing diffusion.
    
The simplest realization of this framework assumes a constant diffusion coefficient. This scenario extends the standard $\Lambda$CDM model by introducing a single additional diffusion parameter and has been extensively analyzed as a background-level modification of the late-time cosmic dynamics~\cite{Calogero:2013zba, Alho:2014ola, Haba:2016swv}. However, there is no a priori fundamental justification for the diffusion strength to remain constant throughout cosmic time. Instead, the diffusion efficiency may realistically depend on evolving physical variables or on the expansion rate of the Universe~\cite{Velten:2014uva, Bandyopadhyay:2019vdd, Corral:2020lxt, LinaresCedeno:2020uxx, Landau:2022mhm}. Motivated by this physical possibility, non-linear diffusion models introduce phenomenological, time-dependent diffusion coefficients. In this work, we systematically investigate four such extensions, wherein the dimensionless diffusion coefficient is parameterized as a function of the scale factor, the matter energy density, the scalar-field density, or the Hubble expansion rate.
	
The primary objective of this paper is to implement and constrain this family of diffusion cosmologies using up-to-date late-time cosmological observations. Specifically, we analyze the baseline constant-diffusion model alongside the four non-linear diffusion parametrizations, comparing their observational performance against both the standard $\Lambda$CDM paradigm and the dynamical Chevallier-Polarski-Linder (CPL) dark energy parametrization~\cite{Calogero:2013zba, Velten:2014uva, Chevallier:2000qy, Linder:2002et}. These models are implemented within the \texttt{SimpleMC} framework~\cite{BOSS:2014hhw, Padilla:2019mgi}, enabling us to compute background-level cosmological observables and perform Bayesian parameter estimation using Type Ia supernova, baryon acoustic oscillation (BAO), and cosmic chronometer Hubble-rate data~\cite{Moresco:2016nqq, Moresco:2020fbm, DESI:2025zgx, Scolnic:2021amr, Brout:2022vxf, Riess:2021jrx}.
	
Our analysis demonstrates that the observational impact of the diffusion mechanism depends strongly on the adopted dataset combination. In the absence of the SH0ES calibration, diffusion models offer a modest improvement in the fit to the data relative to $\Lambda$CDM, performing comparably to the dynamical CPL parametrization. Conversely, when the SH0ES calibration is incorporated, the diffusion models yield a considerably larger reduction in the minimum $\chi^2$, suggesting that diffusion-induced modifications to the late-time expansion history can effectively alleviate the tension with local distance-ladder measurements. However, the statistical advantage of the time-dependent, non-linear diffusion models over the baseline constant-diffusion scenario remains marginal. This indicates that current low-redshift observations are not yet precise enough to robustly discriminate between different functional forms of the diffusion coefficient.
    
The remainder of this paper is organized as follows. In Section~\ref{sec:diffusion_framework}, we introduce the diffusion framework and outline the physical motivation behind both the constant and non-linear diffusion models. Section~\ref{sec:num_imp} presents the resulting background evolution equations and describes the observational datasets used in our analysis, together with their numerical implementation and validation. In Section~\ref{sec:results}, we present the updated parameter constraints and carry out a comprehensive model comparison. Finally, we discuss the physical implications and limitations of our findings and draw our conclusions in Section~\ref{sec:conclusion}.

\section{The Diffusion Framework}\label{sec:diffusion_framework}
	
The matter-diffusion framework provides a general-relativistic description of cosmological models wherein matter particles undergo microscopic velocity diffusion. In this scenario, the matter component is not assumed to be separately conserved; instead, the diffusion process directly modifies the standard matter continuity equation while ensuring that the total energy-momentum tensor remains covariantly conserved. This conservation is achieved by introducing a cosmological scalar field $\phi$ that compensates for the energy exchange induced by diffusion, effectively playing the role of a dynamical dark energy component.
    
Following the relativistic diffusion construction~\cite{Calogero:2011re, Calogero:2012kd, Calogero:2013zba}, the modified gravitational field equations are given by
\begin{equation}\label{sec2_1}
R_{\mu\nu}-\frac{1}{2}g_{\mu\nu}R+\phi g_{\mu\nu}=T_{\mu\nu},
\end{equation}
where $T_{\mu\nu}$ denotes the energy-momentum tensor of the matter component. Here, we adopt geometric units such that $8\pi G=c=1$. The diffusion of matter is governed by the non-conservation equation
\begin{equation}\label{sec2_2}
\nabla_{\mu}T^{\mu\nu}=\sigma J^{\nu},
\end{equation}
supplemented by the conservation of the particle number current,
\begin{equation}\label{sec2_3}
\nabla_{\mu}J^{\mu}=0.
\end{equation}
In these expressions, $J^\mu$ denotes the matter particle four-current, and $\sigma$ is the diffusion coefficient that parametrizes the strength of the microscopic velocity diffusion.
	
The mathematical consistency of this coupled system is guaranteed by the Bianchi identities. Taking the covariant divergence of both sides of the modified Einstein equations~(\ref{sec2_1}) and substituting Eq.~(\ref{sec2_2}) yields the evolution equation for the scalar field:
\begin{equation}\label{sec2_4}
\nabla_{\mu}\phi = \sigma J_{\mu}.
\end{equation}
Therefore, the scalar field evolves in response to the non-conservation of the matter energy-momentum tensor. The matter and scalar-field sectors are not conserved separately, whereas the total system remains covariantly conserved.

For a perfect-fluid matter component, the energy-momentum tensor and particle current are
\begin{equation}\label{sec2_5}
T_{\mu\nu}=\rho u_\mu u_\nu+p(g_{\mu\nu}+u_\mu u_\nu),\qquad
J^\mu=nu^\mu,
\end{equation}
where $\rho$ is the rest-frame energy density, $p$ is the pressure, $n$ is the particle number density, and $u^\mu$ is the fluid four-velocity. In a homogeneous and isotropic FLRW background, the metric is
\begin{equation}\label{sec2_6}
ds^2=-dt^2+a^2(t)\left[\frac{dr^2}{1-Kr^2}+r^2d\Omega^2\right],
\end{equation}
where $K=0,\pm1$ denotes the spatial curvature. The symmetries of the background imply a comoving fluid with $u^\mu=(1,0,0,0)$, and all background quantities depend only on cosmic time.
	
Particle-number conservation then gives
\begin{equation}\label{sec2_7}
n(t)=n_0a^{-3}(t),
\end{equation}
while the matter and scalar-field evolution equations become
\begin{equation}\label{sec2_8}
\dot{\rho}+3H(\rho+p)=\sigma n_0a^{-3},
\end{equation}
and
\begin{equation}\label{sec2_9}
\dot{\phi}=-\sigma n_0a^{-3}.
\end{equation}
The corresponding Friedmann equation is
\begin{equation}\label{sec2_10}
H^2=\frac{1}{3}(\rho+\phi)-\frac{K}{a^2}.
\end{equation}
These equations show that diffusion sources the matter sector, while the scalar field evolves in the opposite direction. In the limit $\sigma=0$, the usual matter conservation equation is recovered and $\phi$ becomes constant, corresponding to the cosmological constant in $\Lambda$CDM.

\subsection{Constant Diffusion}
	
The simplest realization of the diffusion framework is obtained by considering pressureless matter in a spatially flat background. This corresponds to setting
\begin{equation}\label{sec2_11}
p=0, \qquad K=0.
\end{equation}
The resulting model is referred to as the $\phi$CDM model. It consists of a pressureless matter component coupled through diffusion to the scalar field $\phi$, which plays the role of a dynamical dark-energy component.

It is useful to introduce the normalized variables
\begin{equation}\label{sec2_12}
\Omega_m(z)=\frac{\rho_m(z)}{3H_0^2}, \qquad
\Omega_\phi(z)=\frac{\phi(z)}{3H_0^2}, \qquad
E(z)=\frac{H(z)}{H_0},
\end{equation}
together with the dimensionless diffusion parameter
\begin{equation}\label{sec2_13}
\tilde{\sigma}=\frac{\sigma n_0}{3H_0^3}.
\end{equation}
Using the relation between redshift and the scale factor,
\begin{equation}\label{sec2_14}
z=a^{-1}-1,
\end{equation}
the background equations of the constant-diffusion $\phi$CDM model can be written as
\begin{equation}\label{sec2_15}
\frac{d\Omega_m}{dz}=\frac{3\Omega_m}{1+z}-\tilde{\sigma}\,\frac{(1+z)^2}{E(z)},
\end{equation}
\begin{equation}\label{sec2_16}
\frac{d\Omega_\phi}{dz}=\tilde{\sigma}\,\frac{(1+z)^2}{E(z)},
\end{equation}
with
\begin{equation}\label{sec2_17}
E(z)=\sqrt{\Omega_m(z)+\Omega_\phi(z)}.
\end{equation}
	
These equations describe the coupled evolution of the matter and scalar-field sectors. The diffusion term appears with opposite signs in the two evolution equations, reflecting the exchange of energy between the matter and scalar-field sectors. The constant-diffusion model therefore introduces one additional parameter, $\tilde{\sigma}$, relative to the standard spatially flat $\Lambda$CDM model.

As previously discussed, when $\tilde{\sigma}=0$ the diffusion source vanishes. In this limit, the scalar field becomes constant, thereby recovering the standard $\Lambda$CDM solution:
\begin{equation}\label{sec2_18}
\Omega_m(z)=\Omega_{m0}(1+z)^3, \qquad
\Omega_\phi(z)=\Omega_{\phi0}=1-\Omega_{m0}.
\end{equation}
Thus, $\Lambda$CDM is recovered as the zero-diffusion limit of the $\phi$CDM model.

\subsection{Non-linear Diffusion Coefficients}
	
The baseline constant-diffusion model assumes that the diffusion coefficient remains constant throughout cosmic history. However, this assumption is not physically required, as the diffusion strength may naturally depend on cosmological background variables that evolve with the expansion of the Universe. This motivates the exploration of variable, or non-linear, diffusion models.

Phenomenologically, a time-dependent diffusion mechanism can be implemented by promoting the constant dimensionless parameter $\tilde{\sigma}$ appearing in the background evolution equations to an effective function of the cosmological background variables,
\begin{equation}\label{sec2_19}
\tilde{\sigma}\longrightarrow\tilde{\sigma}(z).
\end{equation}
The background equations then become
\begin{equation}\label{sec2_20}
\frac{d\Omega_m}{dz}=\frac{3\Omega_m}{1+z}-\tilde{\sigma}(z)\,\frac{(1+z)^2}{E(z)},
\end{equation}
\begin{equation}\label{sec2_21}
\frac{d\Omega_\phi}{dz}=\tilde{\sigma}(z)\,\frac{(1+z)^2}{E(z)},
\end{equation}
with
\begin{equation}\label{sec2_22}
E(z)=\sqrt{\Omega_m(z)+\Omega_\phi(z)}.
\end{equation}
	
Following the phenomenological formulation of variable matter diffusion, we implement four distinct power-law parameterizations:
\begin{equation}\label{sec2_23}
\tilde{\sigma}_{(n)} \equiv \tilde{\sigma}_0 a^k,
\end{equation}
\begin{equation}\label{sec2_24}
\tilde{\sigma}_{(\rho)} \equiv \tilde{\sigma}_0 \left( \frac{\Omega_m}{\Omega_{m0}} \right)^{\lambda},
\end{equation}
\begin{equation}\label{sec2_25}
\tilde{\sigma}_{(\phi)} \equiv \tilde{\sigma}_0 \left( \frac{\Omega_\phi}{\Omega_{\phi0}} \right)^{\delta},
\end{equation}
and
\begin{equation}\label{sec2_26}
\tilde{\sigma}_{(H)} \equiv \tilde{\sigma}_0 \left( \frac{H}{H_0} \right)^{h_\sigma}.
\end{equation}
Here, $\tilde{\sigma}_0$ represents the present-day diffusion amplitude, while the exponents $k$, $\lambda$, $\delta$, and $h_\sigma$ (not to be confused with the reduced Hubble constant) govern the time dependence of the diffusion coefficient. The first parametrization scales the diffusion coefficient with the scale factor $a$; the second with the normalized matter density; the third with the normalized scalar-field density; and the fourth with the Hubble expansion rate.
	
These parametrizations provide a systematic framework for testing whether cosmological datasets favor a constant diffusion amplitude or one that evolves with cosmic time. In the limit where the power-law index of any given model vanishes ($k,\lambda,\delta,h_\sigma\rightarrow0$), each parametrization reduces identically to the baseline constant-diffusion scenario. Consequently, the constant-diffusion model serves as a nested baseline, facilitating a direct statistical assessment of whether the data justify the introduction of an additional degree of freedom.

\section{Numerical Implementation} \label{sec:num_imp}
	
\subsection{Sampler Code and Datasets}
	
To analyze the diffusion models described in the previous section, we implement them within the \texttt{SimpleMC} parameter-estimation framework~\cite{simplemc}. For each cosmological scenario, the background expansion history is obtained by numerically solving the coupled evolution equations for the matter density and the scalar-field component. The dimensionless Hubble parameter is then given by
\begin{equation}\label{sec3_27}
E(z)\equiv\frac{H(z)}{H_0},
\end{equation}
which serves as the fundamental quantity for computing the late-time observables entering the likelihood functions.

For the baseline constant-diffusion model, the system is evolved using a constant dimensionless diffusion amplitude, $\tilde{\sigma}_0$. For the non-linear extensions, this parameter is replaced by one of the four parametrizations of the diffusion coefficient introduced previously:
\begin{equation}\label{sec3_28}
\tilde{\sigma}_{(n)}, \qquad
\tilde{\sigma}_{(\rho)}, \qquad
\tilde{\sigma}_{(\phi)}, \qquad
\tilde{\sigma}_{(H)}.
\end{equation}
In each configuration, the corresponding diffusion coefficient enters directly into the source term of the coupled background equations. Consequently, a unified numerical implementation is maintained across all diffusion scenarios, with the only difference being the functional form of $\tilde{\sigma}(z)$.
	
The present-day normalization is enforced by imposing
\begin{equation}\label{sec3_29}
E(0)=1,
\end{equation}
subject to the flatness condition
\begin{equation}\label{sec_30}
\Omega_{m0}+\Omega_{\phi0}=1.
\end{equation}
The system is integrated over the redshift range covered by the observational datasets. From the resulting $E(z)$, the code computes the relevant background quantities and cosmological distances, beginning with the Hubble expansion rate,
\begin{equation}\label{sec3_31}
H(z)=H_0E(z),
\end{equation}
and the line-of-sight comoving distance,
\begin{equation}\label{sec3_32}
D_C(z)=\frac{c}{H_0}\int_0^z\frac{dz'}{E(z')}.
\end{equation}
For a spatially flat background, the transverse comoving distance $D_M(z)$ is identical to $D_C(z)$. Consequently, the luminosity distance used to evaluate the Type Ia supernova likelihood is
\begin{equation}\label{sec3_33}
D_L(z)=(1+z)D_C(z).
\end{equation}

In this work, we constrain the diffusion models using combinations of late-time background observations. Since these models modify the homogeneous expansion history, we focus on probes that constrain the late-time expansion history through cosmological distances and measurements of the Hubble expansion rate.

In particular, Type Ia supernova (SN Ia) data constrain the luminosity-distance relation defined in Eq.~(\ref{sec3_33}), thereby providing direct constraints on the late-time expansion history. In the first dataset combination, we use the \texttt{PantheonPlus} likelihood~\cite{Scolnic:2021amr, Brout:2022vxf}, which contains 1701 light curves corresponding to 1550 distinct SNe~Ia and spans the redshift range $0.01<z<2.26$. In the second dataset combination, we use the \texttt{PantheonPlus\&SH0ES} likelihood. This package includes the Pantheon+ supernova sample calibrated using the SH0ES distance ladder~\cite{Riess:2021jrx}, providing tighter constraints on both the absolute distance scale and the Hubble constant, $H_0$.
	
The BAO measurements provide a standard-ruler constraint through the sound horizon at the drag epoch, $r_d$. Depending on the specific BAO measurement, the data constrain different combinations of the transverse comoving distance, the Hubble expansion rate, and the volume-averaged distance. The sampler computes these quantities by evaluating the Hubble distance, $D_H(z)\equiv c/H(z)$, and the spherically averaged volume distance, $D_V(z)$, defined as
\begin{equation}\label{sec3_bao_dv}
\textstyle
D_V(z)\equiv\left[zD_M^2(z)D_H(z)\right]^{1/3}
=\left[\frac{czD_M^2(z)}{H(z)}\right]^{1/3}.
\end{equation}
These distances are then normalized by the sound horizon to yield the dimensionless observables $D_M(z)/r_d$, $D_H(z)/r_d$, and $D_V(z)/r_d$. We use the BAO measurements from the second data release of DESI~\cite{DESI:2025zgx,DESI:2025qqy,DESI:2025fii} and, following~\cite{Schoneberg:2024ifp}, calibrate $r_d$ using Big Bang Nucleosynthesis (BBN) priors. In this work, we use the \texttt{DESI DR2} likelihood as implemented in \texttt{SimpleMC}.

Finally, we include direct measurements of the Hubble expansion rate, denoted as CC, obtained primarily from cosmic chronometers~\cite{Jimenez:2003iv,Simon:2004tf,Moresco:2012jh,Moresco:2012by,Zhang:2012mp,Moresco:2015cya,Moresco:2016mzx,Ratsimbazafy:2017vga,Stern:2009ep}. These measurements constrain $H(z)$ at different redshifts and are particularly valuable for our analysis because the diffusion models modify the background expansion rate directly through the coupled evolution of the matter and scalar-field components.

Therefore, we consider two primary dataset combinations:
\begin{equation}\label{sec4_36}
\mathrm{CC}+\mathrm{DESI}+\mathrm{PantheonPlus},
\end{equation}
and
\begin{equation}\label{sec4_37}
\mathrm{CC}+\mathrm{DESI}+\mathrm{PantheonPlus\&SH0ES}.
\end{equation}
The purpose of these two combinations is to systematically assess the performance of the diffusion models with and without the SH0ES calibration of the Pantheon+ sample~\cite{Riess:2021jrx}.

To establish a baseline for comparison and model selection, we also consider the standard $\Lambda$CDM model and the Chevallier-Polarski-Linder (CPL) parametrization~\cite{Chevallier:2000qy,Linder:2002et}. The $\Lambda$CDM model represents the conventional zero-diffusion limit ($\tilde{\sigma}_0=0$), whereas the CPL model serves as a benchmark for dynamical dark energy characterized by a time-dependent equation of state:
\begin{equation}\label{sec3_34}
w(a)=w_0+w_a(1-a).
\end{equation}
Comparing the diffusion models with these two reference scenarios allows us to determine not only whether the observations favor matter diffusion over a cosmological constant, but also whether it provides a statistical or phenomenological advantage over a conventional dynamical dark energy parametrization with the same number of additional degrees of freedom.
	
In summary, for each dataset combination, we analyze a total of seven models:
\begin{equation}\label{sec3_35}
\Lambda{\rm CDM}, \quad
{\rm CPL}, \quad
\tilde{\sigma}={\rm const.}, \quad
\tilde{\sigma}_{(n)}, \quad
\tilde{\sigma}_{(\rho)}, \quad
\tilde{\sigma}_{(\phi)}, \quad
\tilde{\sigma}_{(H)}.
\end{equation}
The sampled parameters include the standard cosmological parameters and, where applicable, the additional diffusion amplitude and the power-law exponents. The sampled parameters and their corresponding prior ranges are summarized in Table~\ref{parameter_priors}. Unless otherwise stated, all parameters are assigned independent uniform priors.
	
	\begin{table*}[t]
		\centering
		\caption{Uniform prior ranges adopted for the cosmological and model-specific parameters. Here, $\mathcal{U}(x_{\min},x_{\max})$ denotes a uniform prior distribution.}
		\label{parameter_priors}
		
		\small
		\setlength{\tabcolsep}{5pt}
		\renewcommand{\arraystretch}{1.12}
		
		\begin{tabular}{@{} l l c l @{}}
			\toprule
			Parameter
			& \parbox[t]{7.2cm}{Description}
			& Prior
			& Model(s)
			\\
			\midrule
			
			$\Omega_m$
			& \parbox[t]{7.2cm}{Present-day total matter density parameter}
			& $\mathcal{U}(0.10,0.50)$
			& All
			\\
			
			$\Omega_bh^2$
			& \parbox[t]{7.2cm}{Physical baryon density}
			& $\mathcal{U}(0.020,0.025)$
			& All
			\\
			
			$h$
			& \parbox[t]{7.2cm}{Reduced Hubble constant}
			& $\mathcal{U}(0.40,0.90)$
			& All
			\\
			
			$M_B$
			& \parbox[t]{7.2cm}{Type Ia supernova absolute magnitude}
			& $\mathcal{U}(-20.0,-18.5)$
			& SN likelihood
			\\
			
			\midrule
			
			$w_0$
			& \parbox[t]{7.2cm}{Present-day CPL equation-of-state parameter}
			& $\mathcal{U}(-2.0,0.0)$
			& CPL
			\\
			
			$w_a$
			& \parbox[t]{7.2cm}{CPL evolution parameter}
			& $\mathcal{U}(-4.0,2.0)$
			& CPL
			\\
			
			\midrule
			
			$\tilde{\sigma}_0$
			& \parbox[t]{7.2cm}{Present-day dimensionless diffusion amplitude}
			& $\mathcal{U}(0.0,0.4)$
			& All diffusion models
			\\
			
			$k$
			& \parbox[t]{7.2cm}{Scale-factor diffusion exponent}
			& $\mathcal{U}(-5.0,5.0)$
			& $\tilde{\sigma}_{(n)}$
			\\
			
			$\lambda$
			& \parbox[t]{7.2cm}{Matter-density diffusion exponent}
			& $\mathcal{U}(-5.0,5.0)$
			& $\tilde{\sigma}_{(\rho)}$
			\\
			
			$\delta$
			& \parbox[t]{7.2cm}{Scalar-field-density diffusion exponent}
			& $\mathcal{U}(-5.0,5.0)$
			& $\tilde{\sigma}_{(\phi)}$
			\\
			
			$h_\sigma$
			& \parbox[t]{7.2cm}{Expansion-rate diffusion exponent}
			& $\mathcal{U}(-5.0,5.0)$
			& $\tilde{\sigma}_{(H)}$
			\\
			
			\bottomrule
		\end{tabular}
	\end{table*}

\subsection{Validation Against Reference Results}
	
Before deriving cosmological constraints, we validate our numerical implementation by comparing the reconstructed background evolution with the reference behavior of the variable diffusion models. This step is important because the diffusion coefficient enters directly into the coupled evolution equations for the matter and scalar-field components. Therefore, an incorrect implementation would immediately affect the predicted expansion history.

\begin{figure*}[t]
	\centering
	
	\begin{minipage}[b]{0.45\textwidth}
		\centering
		\includegraphics[width=\linewidth]{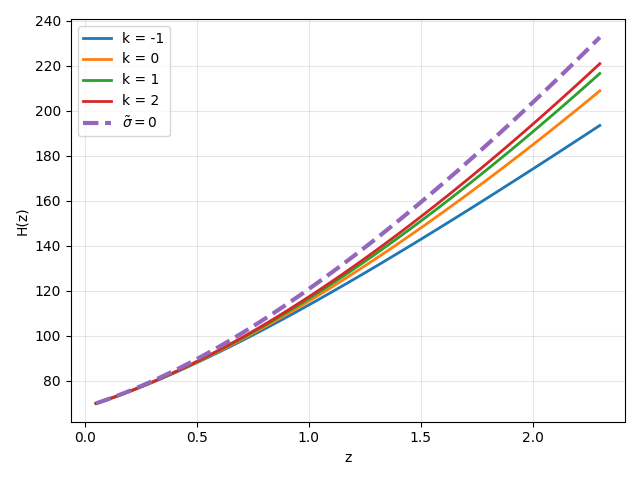}
	\end{minipage}
	\hfill
	\begin{minipage}[b]{0.45\textwidth}
		\centering
		\includegraphics[width=\linewidth]{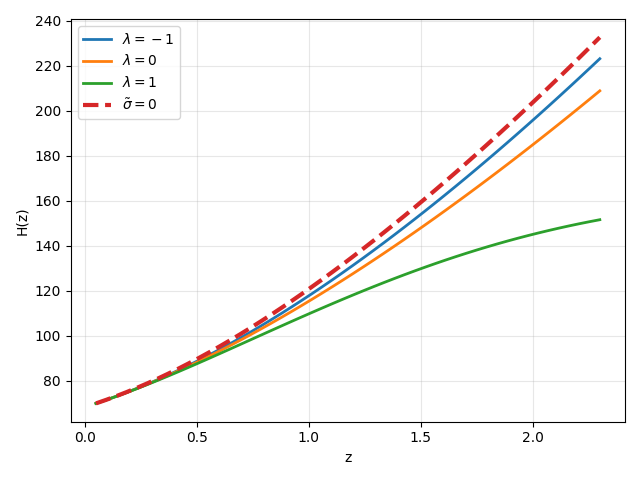}
	\end{minipage}
	
	\vspace{0.3cm}
	
	\begin{minipage}[b]{0.45\textwidth}
		\centering
		\includegraphics[width=\linewidth]{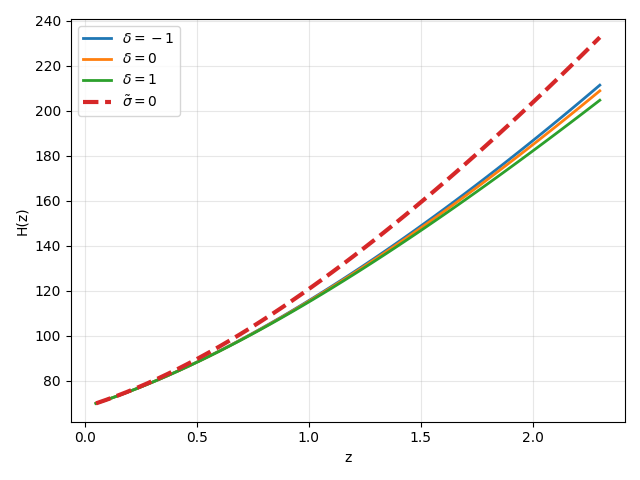}
	\end{minipage}
	\hfill
	\begin{minipage}[b]{0.45\textwidth}
		\centering
		\includegraphics[width=\linewidth]{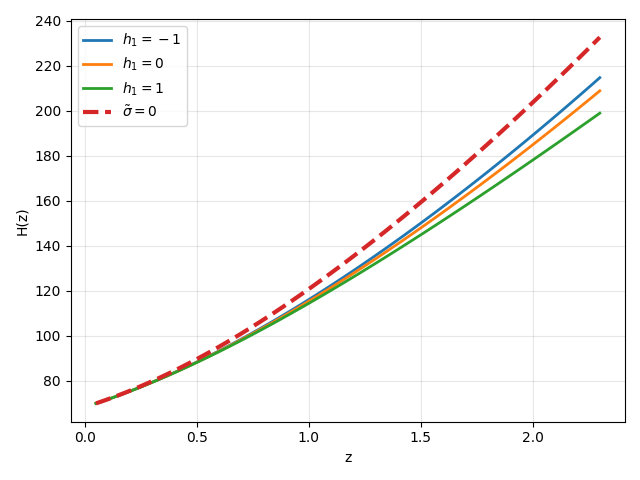}
	\end{minipage}
	
\caption{Hubble expansion rate, $H(z)$, obtained from the \texttt{SimpleMC}
implementation for the four variable matter-diffusion
parametrizations. The upper-left panel corresponds to the
scale-factor-dependent model,
$\tilde{\sigma}_{(n)}=\tilde{\sigma}_0a^k$; the upper-right
panel to the matter-density-dependent model,
$\tilde{\sigma}_{(\rho)}
=\tilde{\sigma}_0
(\Omega_m/\Omega_{m0})^\lambda$; the lower-left
panel to the scalar-field-dependent model,
$\tilde{\sigma}_{(\phi)}
=\tilde{\sigma}_0
(\Omega_\phi/\Omega_{\phi0})^\delta$; and the lower-right
panel to the Hubble-rate-dependent model,
$\tilde{\sigma}_{(H)}
=\tilde{\sigma}_0(H/H_0)^{h_\sigma}$.
The dashed curve in each panel represents the zero-diffusion
limit, $\tilde{\sigma}=0$, corresponding to the
$\Lambda$CDM background. The solid curves illustrate the effect of varying the corresponding power-law exponent governing the evolution of the diffusion coefficient.
}
	\label{Hz_validation}
\end{figure*}

Figure~\ref{Hz_validation} shows the Hubble expansion rate, $H(z)$, obtained from our \texttt{SimpleMC} implementation for the four variable diffusion parametrizations. Each panel corresponds to one functional dependence of the diffusion coefficient: the scale factor, the matter density, the scalar-field density, and the Hubble expansion rate. The dashed curve represents the zero-diffusion limit, $\tilde{\sigma}=0$, corresponding to the standard $\Lambda$CDM background. The comparison illustrates how varying the exponent of each parametrization modifies the late-time expansion history.
	
\begin{figure*}[t]
		\centering
		\includegraphics[width=0.92\textwidth]{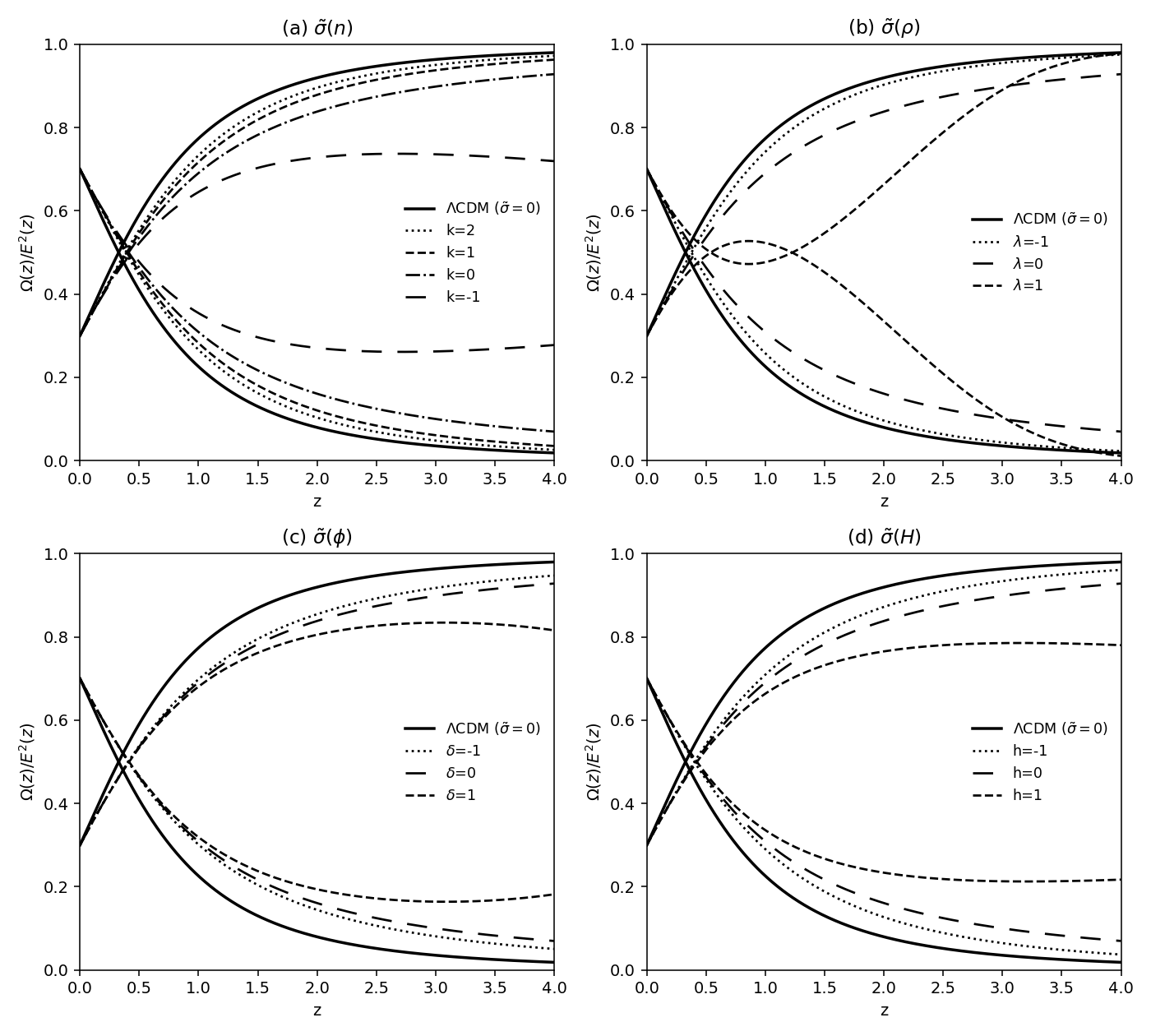}
		\caption{
Background evolution reproduced for the variable matter-diffusion models. Each panel corresponds to one diffusion parametrization: $\tilde{\sigma}_{(n)}$, $\tilde{\sigma}_{(\rho)}$, $\tilde{\sigma}_{(\phi)}$, and $\tilde{\sigma}_{(H)}$. The curves show the fractional energy densities, $\Omega_i(z)/E^2(z)$, for the matter ($i=m$) and scalar-field ($i=\phi$) components. The solid curves correspond to the zero-diffusion ($\Lambda$CDM) limit, while the remaining line styles illustrate different choices of the power-law exponent governing the time dependence of the diffusion coefficient.
}
		\label{density_validation}
	\end{figure*}

Figure~\ref{density_validation} presents the reproduced background evolution of the normalized density fractions for the same four classes of variable diffusion models. The plotted quantities are the fractional contributions $\Omega_i(z)/E^2(z)$, where $i=m,\phi$. As expected, the matter contribution increases toward high redshift, while the scalar-field contribution decreases, reflecting the transition from matter domination to late-time dark-energy domination. The different parametrizations modify the detailed evolution of the two components while preserving the expected asymptotic behavior. The excellent agreement with the reference results confirms the correct numerical implementation of the coupled diffusion equations.
	
This validation confirms that the numerical implementation is stable and that the diffusion source terms are correctly implemented. The validated pipeline is then used for the cosmological parameter estimation and model comparison presented in the following sections.

\section{Results}\label{sec:results}

\begin{figure}[]
	\centering
	\includegraphics[
	width=0.45\textwidth
	]{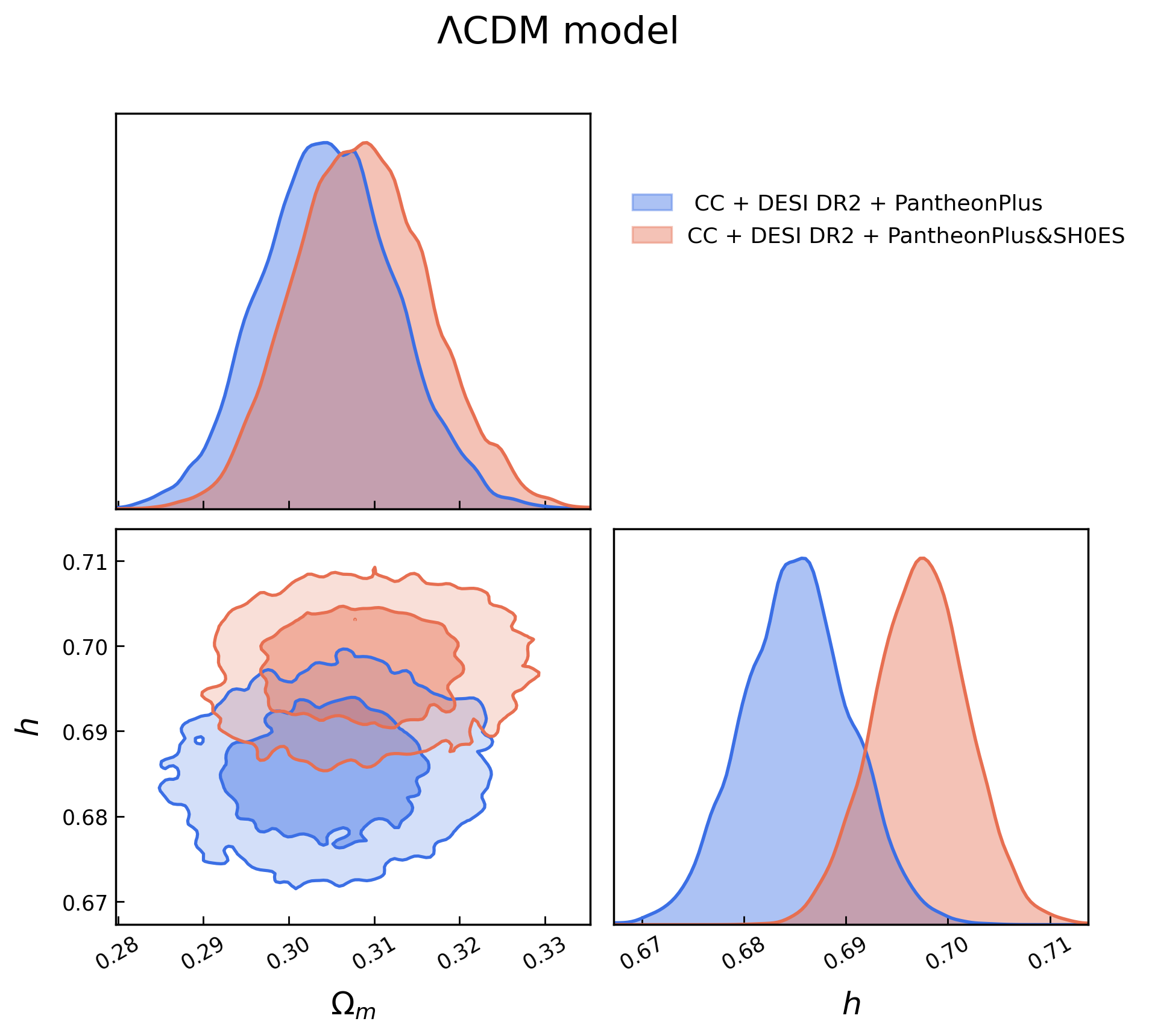}
	
	\caption{Marginalized posterior distributions for the $\Lambda$CDM parameter set $\{\Omega_m,h\}$. Including the SH0ES-calibrated dataset shifts the preferred value of $h$ toward larger values, accompanied by a smaller correlated increase in $\Omega_m$.}
	\label{fig_lcdm_corner}
\end{figure}

\begin{figure}[!t]
	\centering
	\includegraphics[
	width=0.5\textwidth
	]{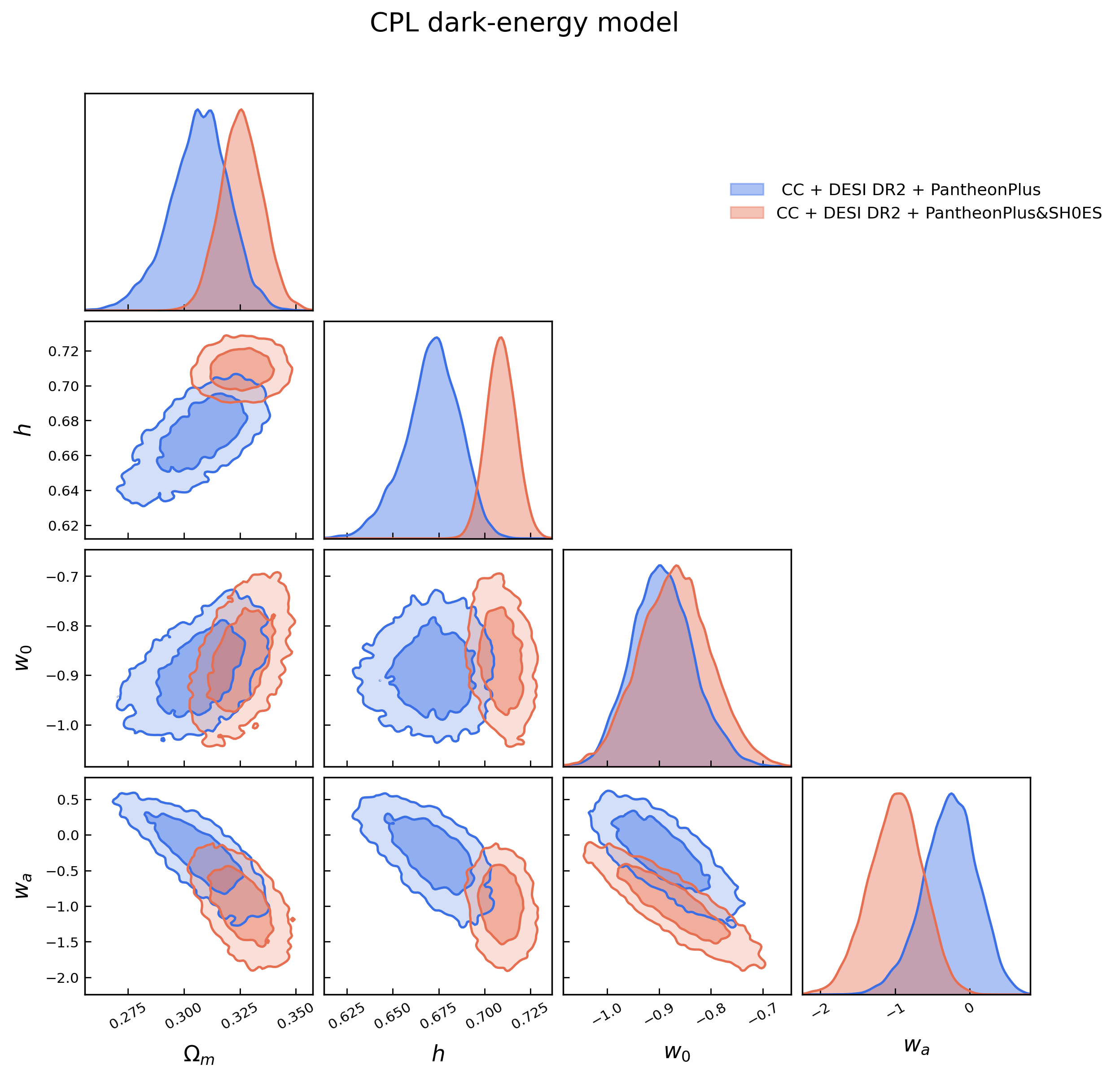}
	
	\caption{
Marginalized posterior distributions for the CPL parameter set
$\{\Omega_m,h,w_0,w_a\}$. The characteristic anti-correlation between
$w_0$ and $w_a$ is clearly visible, together with their degeneracies with
$\Omega_m$ and $h$. Including the SH0ES-calibrated dataset shifts the posterior only moderately, leaving the overall parameter constraints broadly unchanged.
}
	\label{fig_cpl_corner}
\end{figure}

\begin{figure}[!t]
	\centering
	\includegraphics[
	width=0.5\textwidth
	]{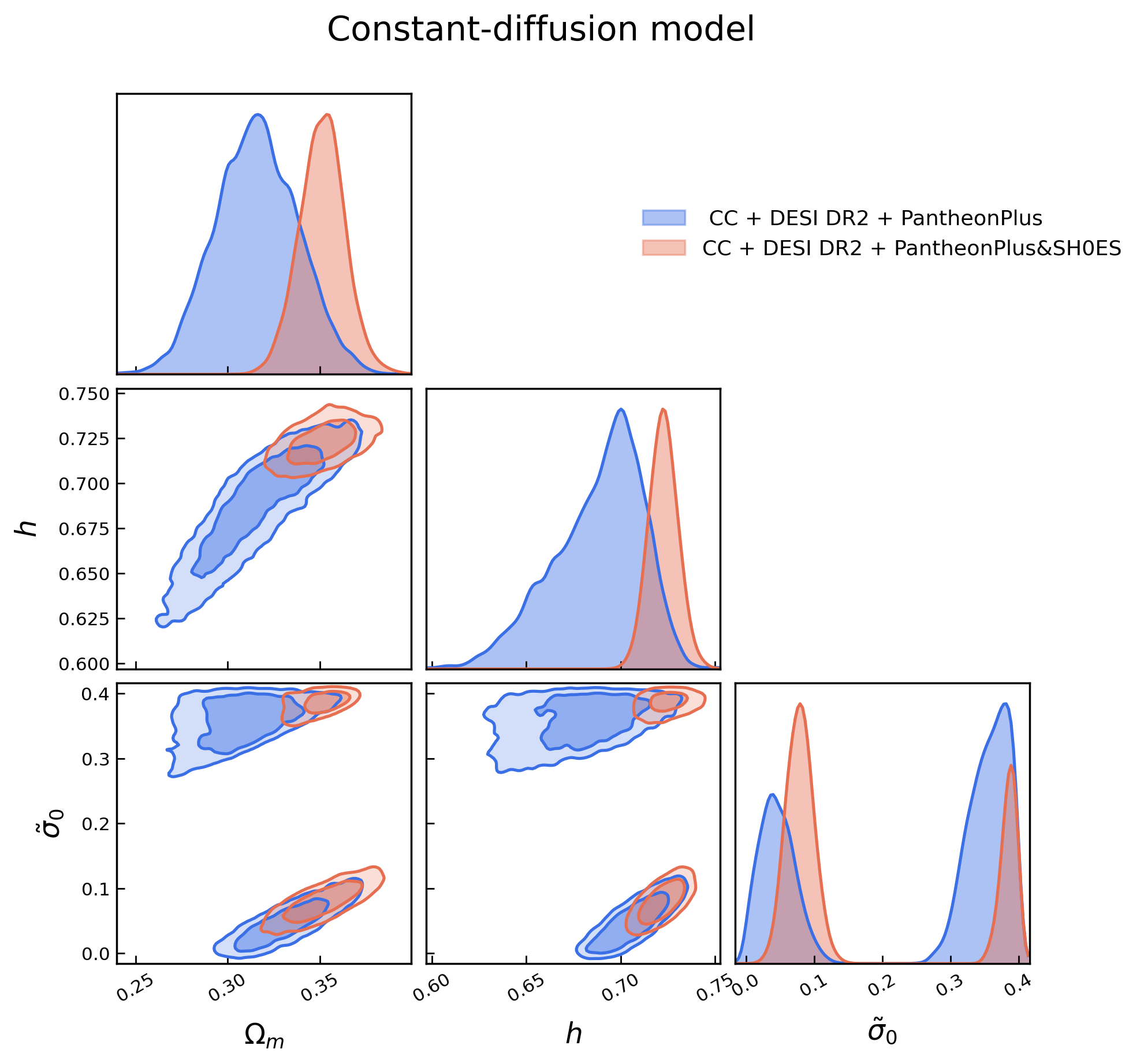}
	
\caption{
Marginalized posterior distributions for the constant-diffusion
parameter set $\{\Omega_m,h,\tilde{\sigma}_0\}$. The posterior of the
diffusion amplitude is clearly bimodal and exhibits significant
degeneracies with both $\Omega_m$ and $h$. Including the
SH0ES-calibrated dataset shifts the relative statistical weight and
location of the two preferred diffusion regimes while preserving the
overall bimodal structure.
}
	\label{fig_constant_corner}
\end{figure}

\begin{figure}[!t]
	\centering
	\includegraphics[
	width=0.5\textwidth
	]{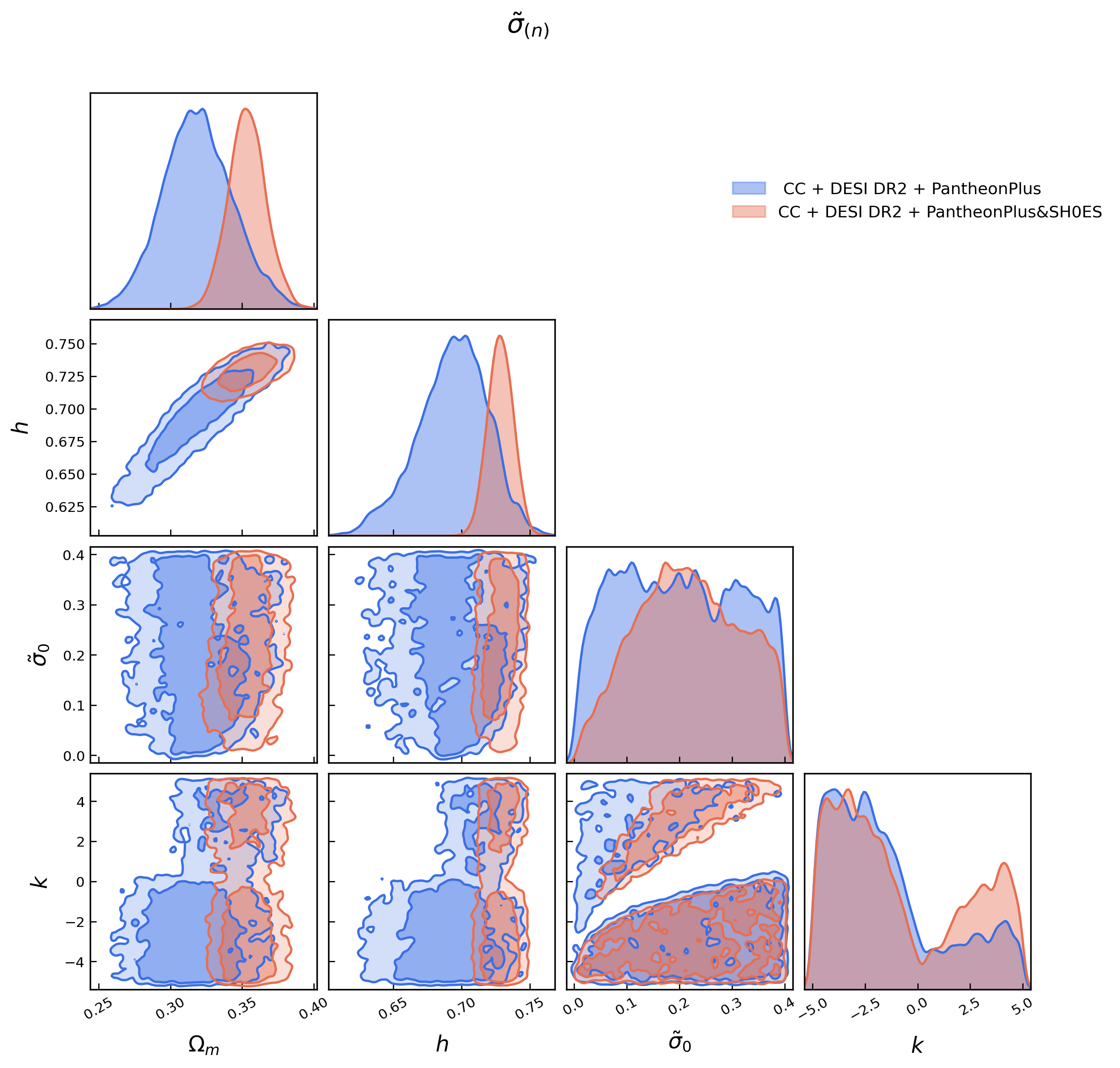}
	
\caption{
Marginalized posterior distributions for the scale-factor-dependent
diffusion model,
$\tilde{\sigma}_{(n)}=\tilde{\sigma}_0a^k$, with parameter set
$\{\Omega_m,h,\tilde{\sigma}_0,k\}$. The broad posterior in the
$\tilde{\sigma}_0$--$k$ plane illustrates the degeneracy between the
present-day diffusion amplitude and its scale-factor dependence. The
posterior of $k$ exhibits a mild bimodal structure while remaining
compatible with the constant-diffusion limit ($k=0$).
}
	\label{fig_sigma_n_corner}
\end{figure}

\begin{figure}[!t]
	\centering
	\includegraphics[
	width=0.5\textwidth
	]{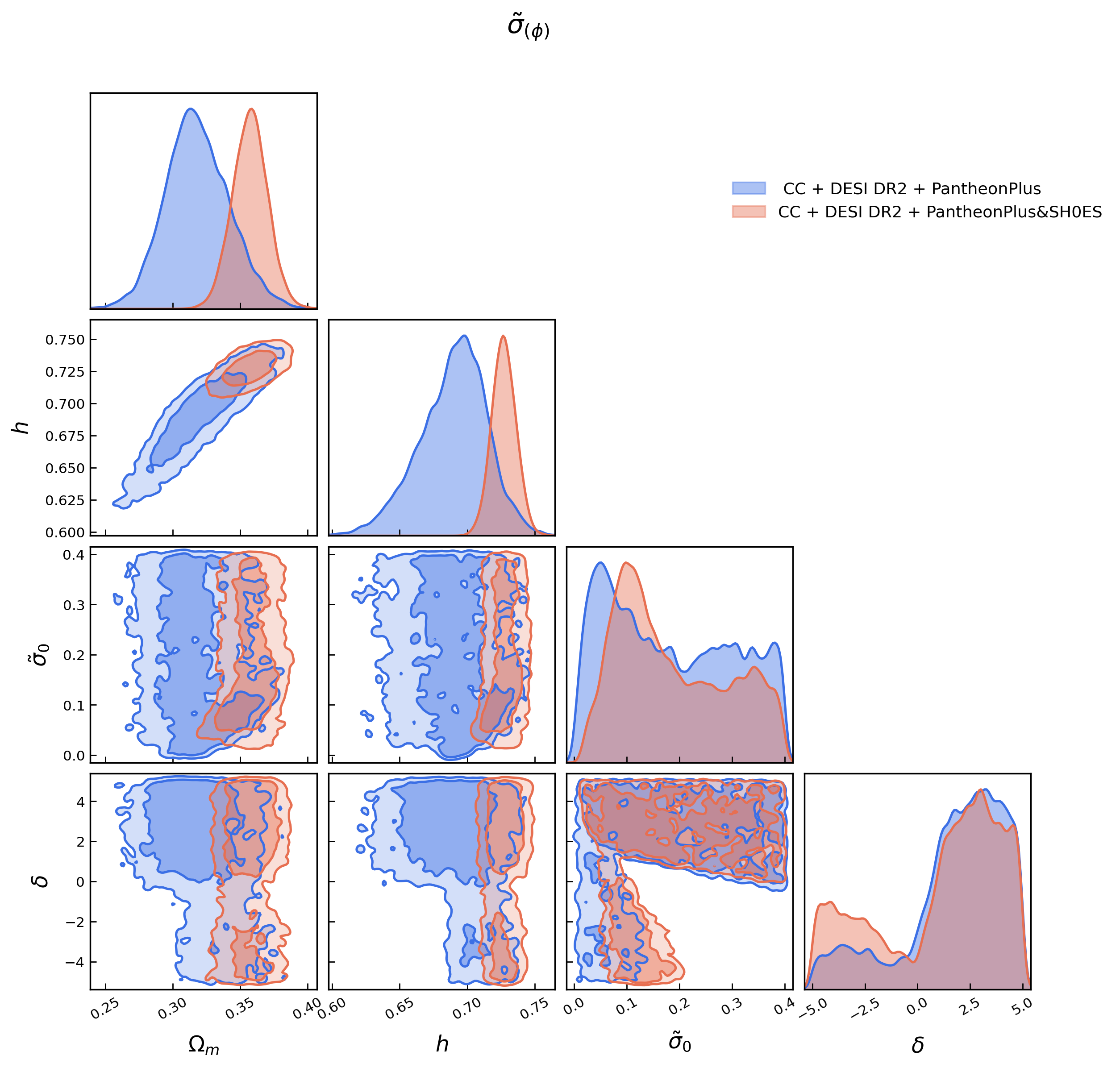}
	
\caption{
Marginalized posterior distributions for the scalar-field-dependent
diffusion model,
$\tilde{\sigma}_{(\phi)}
=\tilde{\sigma}_0(\Omega_\phi/\Omega_{\phi0})^\delta$,
with parameter set
$\{\Omega_m,h,\tilde{\sigma}_0,\delta\}$. The broad posterior of
$\delta$ reflects the limited sensitivity of the data to the detailed
scalar-field dependence of the diffusion coefficient. A weaker bimodal
structure in $\tilde{\sigma}_0$ is still apparent, while the posterior
of $\delta$ remains compatible with the constant-diffusion limit
($\delta=0$) and exhibits only a mild hint of bimodality.
}
	\label{fig_sigma_phi_corner}
\end{figure}

\begin{figure}[]
	\centering
	\includegraphics[
	width=0.5\textwidth
	]{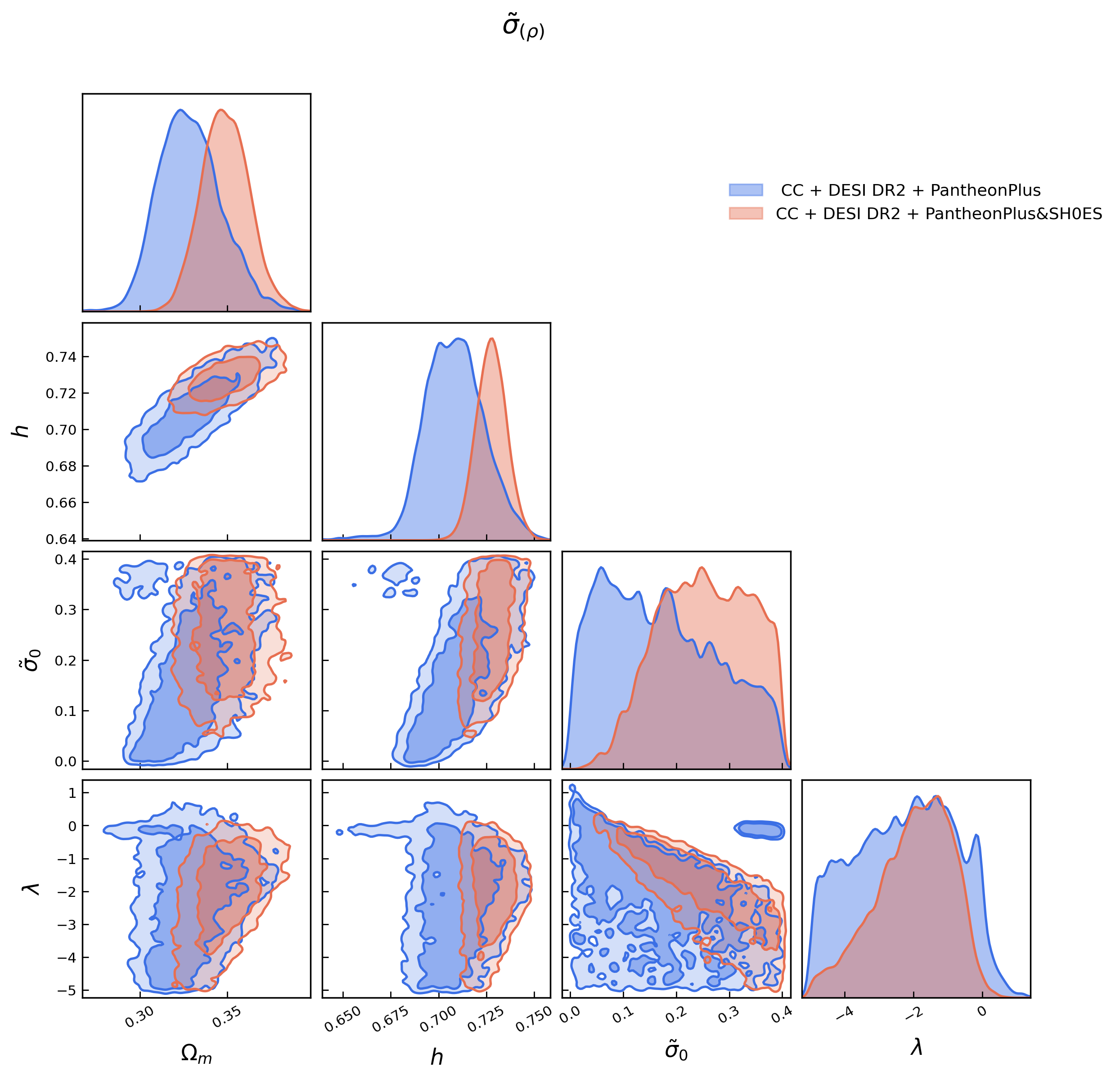}
	
\caption{
Marginalized posterior distributions for the matter-density-dependent
diffusion model,
$\tilde{\sigma}_{(\rho)}
=\tilde{\sigma}_0(\Omega_m/\Omega_{m0})^\lambda$,
with parameter set
$\{\Omega_m,h,\tilde{\sigma}_0,\lambda\}$. The contours reveal a
degeneracy between $\lambda$ and the diffusion amplitude, together with
correlations with the background cosmological parameters. The posterior
is confined to negative values of $\lambda$, corresponding to a
diffusion coefficient that is suppressed at higher redshift and becomes
more relevant at late times.
}
	\label{fig_sigma_rho_corner}
\end{figure}

\begin{figure}[]
	\centering
	\includegraphics[
	width=0.5\textwidth
	]{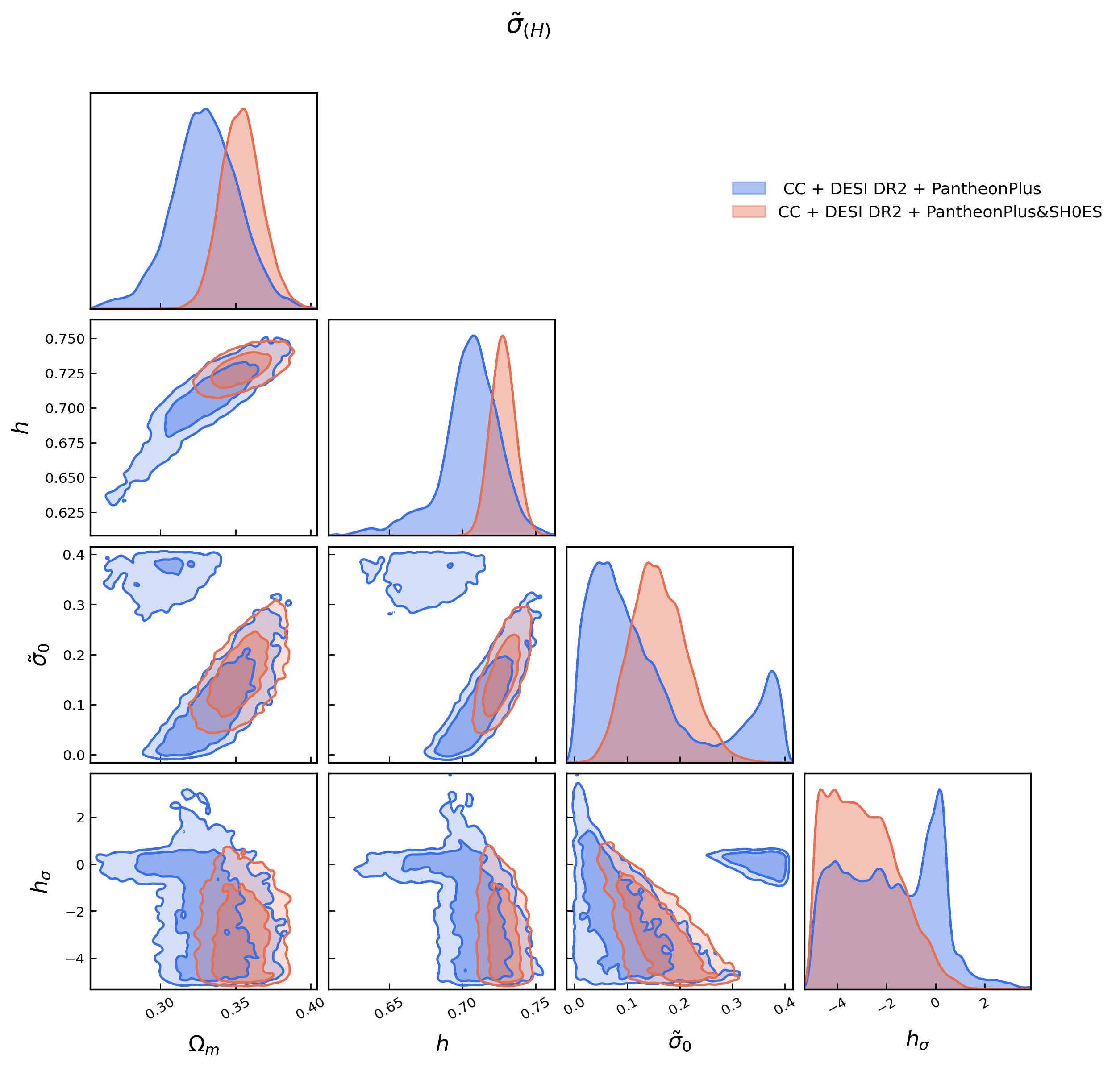}
	
\caption{
Marginalized posterior distributions for the Hubble-rate-dependent
diffusion model,
$\tilde{\sigma}_{(H)}
=\tilde{\sigma}_0(H/H_0)^{h_\sigma}$,
with parameter set
$\{\Omega_m,h,\tilde{\sigma}_0,h_\sigma\}$. The exponent $h_\sigma$
is correlated with the diffusion amplitude, and a bimodal structure
in $\tilde{\sigma}_0$ is again visible. While both positive and negative
values of $h_\sigma$ are allowed, the SH0ES-calibrated data shift the
posterior weight toward the negative branch. Since $H/H_0$ increases
with redshift, negative values of $h_\sigma$ suppress the diffusion
coefficient at earlier times and enhance its relative impact at late
times.
}
	\label{fig_sigma_h_corner}
\end{figure}

In this section, we present the cosmological parameter constraints obtained for the seven models under consideration using the two late-time dataset combinations. The marginalized 68\% confidence intervals and best-fit values are summarized in Tables~\ref{tab:constraints_fastpantheon_DESIBAO_DR2_hd} and~\ref{tab:evidence_fastpantheon}, while the corresponding posterior distributions are shown in Figures~\ref{fig_lcdm_corner},
\ref{fig_cpl_corner},
\ref{fig_constant_corner},
\ref{fig_sigma_n_corner},
\ref{fig_sigma_phi_corner},
\ref{fig_sigma_rho_corner}, and
\ref{fig_sigma_h_corner}. Throughout the corner plots, the blue contours correspond to the baseline dataset combination without the SH0ES calibration, whereas the red contours illustrate the impact of including it.

Furthermore, the nested-sampling framework enables a rigorous model comparison through the computation of the Bayesian evidence, or marginal likelihood, $\mathcal{Z}$. We define the relative log-evidence as
\begin{equation}\label{eq:delta_lnZ}
\Delta\ln\mathcal{Z}_i \equiv \ln\mathcal{Z}_i-\ln\mathcal{Z}_{\Lambda\mathrm{CDM}},
\end{equation}
where $\Delta\ln\mathcal{Z}_i>0$ indicates a Bayesian preference for model $i$ over the $\Lambda$CDM baseline. Following the revised Jeffreys scale~\cite{Trotta:2008qt}, the evidence is interpreted as inconclusive if $|\Delta\ln\mathcal{Z}_i|<1$, weak if $1<|\Delta\ln\mathcal{Z}_i|<2.5$, moderate if $2.5<|\Delta\ln\mathcal{Z}_i|<5$, and strong if $|\Delta\ln\mathcal{Z}_i|>5$. For direct comparison with the maximum-likelihood improvement, we similarly define
\begin{equation}\label{eq:delta_chi2}
\Delta\chi^2_i \equiv \chi^2_i-\chi^2_{\Lambda\mathrm{CDM}},
\end{equation}
where negative values ($\Delta\chi^2_i<0$) indicate an improved fit to the data.


\subsection{Parameter Constraints and goodness of the fit}
	
The cosmological parameter constraints derived from the dataset combination without the local distance-ladder calibration ($\mathrm{CC}+\mathrm{DESI\,DR2}+\mathrm{PantheonPlus}$) are summarized in Table~\ref{tab:constraints_fastpantheon_DESIBAO_DR2_hd}. Compared to $\Lambda$CDM, the diffusion models exhibit significantly broader posterior distributions for both $\Omega_m$ and $H_0$, reflecting the additional parameter degeneracies introduced by the diffusion sector. In particular, the uncertainty on $H_0$ increases by up to a factor of five relative to $\Lambda$CDM, while that on $\Omega_m$ increases by approximately a factor of three. Relative to the CPL parametrization, the uncertainties remain approximately a factor of two larger. Despite these enlarged error bars, the preferred values display a mild shift toward larger matter densities, from $\Omega_m\simeq0.30$ in $\Lambda$CDM to $\Omega_m\simeq0.31$ for CPL and $\Omega_m\simeq0.32$--$0.33$ for the diffusion models. A similar trend is observed for the Hubble constant. While the CPL parametrization slightly lowers the preferred value relative to $\Lambda$CDM, the diffusion models generally recover values comparable to the $\Lambda$CDM solution, with the $\rho$-dependent model exhibiting the highest preferred value, $h\simeq0.71$. Although these shifts are not statistically significant because of the substantially larger uncertainties, they naturally move the preferred parameter space toward better agreement with the local SH0ES determination discussed below.

Within the diffusion framework, the constraints on the present-day diffusion amplitude depend on the adopted parametrization. In particular, the constant-diffusion model exhibits a strongly bimodal posterior for $\tilde{\sigma}_0$, making a single posterior mean an inadequate summary of the constraint. For the $\mathrm{CC}+\mathrm{DESI\,DR2}+\mathrm{PantheonPlus}$ dataset combination, the two preferred branches are centered at approximately $\tilde{\sigma}_0\simeq0.044$ and $0.360$. This bimodality can be understood qualitatively from the coupled background evolution: different diffusion strengths modify the relative contributions of the matter and scalar-field sectors, while the expansion rate depends only on their sum, $E^2(z)=\Omega_m(z)+\Omega_\phi(z)$. Consequently, distinct diffusion regimes can yield comparable background expansion histories over the redshift range probed by the data. 

In contrast, the non-linear models consistently favor intermediate values around $\tilde{\sigma}_0\simeq0.2$, with comparable uncertainties.
The additional power-law exponents remain broadly compatible with the constant-diffusion limit, corresponding to a vanishing exponent. The largest departure is found for the matter-density-dependent model, with $\lambda=-2.3\pm1.4$, corresponding to a mild $\sim1.6\sigma$ preference for negative values. Therefore, the current late-time dataset combination does not provide statistically significant evidence for a time-dependent diffusion coefficient.

For this baseline dataset combination, the standard $\Lambda$CDM model yields a minimum $\chi^2 = 1431.44$. Introducing a dynamical dark-energy equation of state through the two-parameter CPL parametrization reduces this to $\chi^2 = 1427.36$, corresponding to an improvement of $\Delta\chi^2 = -4.08$ relative to $\Lambda$CDM.
The constant-diffusion model provides a comparable fit, yielding $\chi^2 = 1427.78$ ($\Delta\chi^2 = -3.66$). The four non-linear diffusion extensions provide only marginal additional improvements, with minimum $\chi^2$ values clustering around $\sim1427$. The lowest value within this family is obtained for the Hubble-dependent diffusion model, $\tilde{\sigma}_{(H)}$, with $\chi^2 = 1427.25$ ($\Delta\chi^2 = -4.19$ relative to $\Lambda$CDM). However, its improvement over the single-parameter constant-diffusion model is only $\Delta\chi^2 = -0.53$, indicating that the current late-time dataset combination does not statistically justify the additional freedom introduced by a time-dependent diffusion coefficient.

\begin{table*}[t]
	\centering
	\caption{
Marginalized parameter constraints for
$\mathrm{CC}+\mathrm{DESI\,DR2}+\mathrm{PantheonPlus}$.
The last column reports the minimum $\chi^2$.
}
	\label{tab:constraints_fastpantheon_DESIBAO_DR2_hd}
	
	\small
	\setlength{\tabcolsep}{4.5pt}
	\renewcommand{\arraystretch}{1.12}
	
	\begin{tabular*}{\textwidth}{
			@{\extracolsep{\fill}}
			lccccccc
		}
		\toprule
		Model
		& $\Omega_m$
		& $h$
		& $w_0$
		& $w_a$
		& $\tilde{\sigma}_0$
		& Exponent
		& $\chi^2_{\min}$
		\\
		\midrule
		
		$\Lambda$CDM
		& $0.30 \pm 0.008$
		& $0.69 \pm 0.005$
		& --
		& --
		& --
		& --
		& $1431.44$
		\\
		
		CPL
		& $0.31 \pm 0.013$
		& $0.67 \pm 0.015$
		& $-0.89 \pm 0.059$
		& $-0.27 \pm 0.36$
		& --
		& --
		& $1427.36$
		\\
		
		Constant
		& $0.32 \pm 0.022$
		& $0.69 \pm 0.024$
		& --
		& --
        & $\begin{array}{c}
        0.044^{+0.028}_{-0.024} \\
        0.360^{+0.027}_{-0.034}
        \end{array}$
		& --
		& $1427.78$
		\\
		
		$\phi$-power $(\delta)$
		& $0.32 \pm 0.024$
		& $0.69 \pm 0.025$
		& --
		& --
		& $0.18 \pm 0.12$
		& $\delta=1.8 \pm 2.4$
		& $1427.40$
		\\
		
		$n$-power $(k)$
		& $0.32 \pm 0.024$
		& $0.69 \pm 0.026$
		& --
		& --
		& $0.20 \pm 0.11$
		& $k=-1.5 \pm 2.7$
		& $1427.31$
		\\
		
		$\rho$-power $(\lambda)$
		& $0.33 \pm 0.018$
		& $0.71 \pm 0.016$
		& --
		& --
		& $0.17 \pm 0.11$
		& $\lambda=-2.3 \pm 1.4$
		& $1427.34$
		\\
		
		$H$-power $(h_\sigma)$
		& $0.33 \pm 0.022$
		& $0.70 \pm 0.021$
		& --
		& --
		& $0.14 \pm 0.12$
		& $h_\sigma=-1.9 \pm 1.9$
		& $1427.25$
		\\
		
		\bottomrule
	\end{tabular*}
\end{table*}

The resulting Bayesian model comparison metrics are presented in Table~\ref{tab:evidence_fastpantheon}. Although all extended frameworks improve the best-fit $\chi^2$ by $\Delta\chi^2 \approx -3.7$ to $-4.2$, this improvement is not accompanied by a Bayesian preference over $\Lambda$CDM. For the diffusion model family, the evidence differences fall within the narrow range
\begin{equation}\label{eq:lnZ_range_fastpantheon}
-0.74 \lesssim \Delta\ln\mathcal{Z} \lesssim -0.14.
\end{equation}
According to the Jeffreys scale, values of $|\Delta\ln\mathcal{Z}| < 1$ correspond to inconclusive evidence. Therefore, the diffusion models remain statistically indistinguishable from $\Lambda$CDM for this dataset combination, indicating that the modest improvement in the maximum likelihood is offset by the Occam penalty associated with the larger prior volume. In contrast, the CPL parametrization incurs a substantially larger Bayesian penalty, yielding $\Delta\ln\mathcal{Z} = -2.45 \pm 0.19$, indicating moderate evidence against CPL relative to $\Lambda$CDM. This reflects the fact that its two additional parameters ($w_0$, $w_a$) increase the prior volume without providing a substantially better fit than either $\Lambda$CDM or the single-parameter constant-diffusion model.

\begin{table*}[t]
	\centering
\caption{
Bayesian model comparison for
$\mathrm{CC}+\mathrm{DESI\,DR2}+\mathrm{PantheonPlus}$.
All quantities are reported relative to $\Lambda$CDM.
}
	\label{tab:evidence_fastpantheon}
	
	\setlength{\tabcolsep}{10pt}
	\renewcommand{\arraystretch}{1.12}
	
	\begin{tabular}{lcccc}
		\toprule
		Model
		& $\chi^2_{\min}$
		& $\Delta\chi^2$
		& $\ln\mathcal{Z}$
		& $\Delta\ln\mathcal{Z}$
		\\
		\midrule
		
		$\Lambda$CDM
		& $1431.44$
		& $0.00$
		& $-724.199 \pm 0.118$
		& $0.000$
		\\
		
		CPL
		& $1427.36$
		& $-4.08$
		& $-726.651 \pm 0.146$
		& $-2.451 \pm 0.188$
		\\
		
		Constant diffusion
		& $1427.78$
		& $-3.66$
		& $-724.941 \pm 0.132$
		& $-0.742 \pm 0.178$
		\\
		
		$\phi$-power $(\delta)$
		& $1427.40$
		& $-4.04$
		& $-724.862 \pm 0.132$
		& $-0.663 \pm 0.177$
		\\
		
		$n$-power $(k)$
		& $1427.31$
		& $-4.13$
		& $-724.527 \pm 0.129$
		& $-0.328 \pm 0.175$
		\\
		
		$\rho$-power $(\lambda)$
		& $1427.34$
		& $-4.10$
		& $-724.461 \pm 0.126$
		& $-0.262 \pm 0.173$
		\\
		
		$H$-power $(h_\sigma)$
		& $1427.25$
		& $-4.19$
		& $-724.341 \pm 0.128$
		& $-0.142 \pm 0.174$
		\\
		
		\bottomrule
	\end{tabular}
\end{table*}

Conversely, when the local SH0ES calibration is incorporated, the impact of the diffusion sector becomes considerably more pronounced, as summarized in Table~\ref{tab:constraints_pantheonplus_shoes}. While the $\Lambda$CDM and CPL models remain unable to accommodate the higher Hubble constant preferred by the local distance ladder, all diffusion scenarios shift toward values fully consistent with the SH0ES determination. This upward shift in $H_0$ is accompanied by a corresponding increase in the preferred matter density, with $\Omega_m$ increasing from approximately $0.30$ in $\Lambda$CDM to about $0.35$ in the diffusion models. The enlarged posterior distributions observed for the diffusion scenarios naturally allow the models to explore this region of parameter space while remaining compatible with the late-time observations.

The behavior of the diffusion parameters also changes after including the SH0ES calibration. The constant-diffusion model continues to exhibit a pronounced bimodal posterior, although the two preferred branches shift to approximately $\tilde{\sigma}_0\simeq0.079$ and $0.387$. The SH0ES calibration therefore changes both the location and the relative statistical weight of the two allowed diffusion regimes, rather than selecting a unique value of the diffusion amplitude. For the non-linear models, the preferred value remains close to $\tilde{\sigma}_0\simeq0.2$, with comparable uncertainties.
The strongest departures from the constant-diffusion limit are now obtained for the matter-density-dependent and Hubble-dependent parametrizations, with $\lambda=-2.0\pm1.1$ and $h_{\sigma}=-2.9\pm1.4$, corresponding to approximately $1.8\sigma$ and $2.1\sigma$ departures from zero, respectively, while $k$ and $\delta$ remain only weakly constrained. The tendency toward negative values of $\lambda$ and $h_{\sigma}$ has a simple physical interpretation: since both $\Omega_m/\Omega_{m0}$ and $H/H_0$ increase toward higher redshift, negative exponents suppress the diffusion coefficient in the past and enhance its effect at late times. This behavior is qualitatively consistent with the stronger preference for diffusion obtained when the SH0ES calibration is included. Nevertheless, the statistical significance remains insufficient to establish compelling evidence for a specific time-dependent diffusion law.

For this dataset combination, the standard $\Lambda$CDM baseline yields a minimum $\chi^2_{\mathrm{min}} = 1501.90$. Introducing dynamical dark energy through the two-parameter CPL parametrization reduces this value to $\chi^2_{\mathrm{min}} = 1491.03$, corresponding to an improvement of $\Delta\chi^2 = -10.87$ relative to $\Lambda$CDM.
The single-parameter constant-diffusion model provides a substantially larger improvement, reaching $\chi^2_{\mathrm{min}} = 1480.50$, corresponding to $\Delta\chi^2 = -21.40$ relative to $\Lambda$CDM and $\Delta\chi^2 = -10.53$ relative to CPL. This indicates that matter diffusion accommodates the local distance-ladder calibration considerably more effectively than the standard CPL dynamical dark-energy parametrization.

In contrast, the non-linear diffusion extensions yield only marginal additional reductions in $\chi^2_{\mathrm{min}}$. The lowest value is obtained for the Hubble-dependent model, $\tilde{\sigma}_{(H)}$, with $\Delta\chi^2 = -22.48$. However, its improvement over the constant-diffusion model is only $\Delta\chi^2 = -1.08$, indicating that the additional freedom associated with a time-dependent diffusion coefficient is not statistically justified by the current data.

\begin{table*}[t]
	\centering
	\caption{
Marginalized parameter constraints for
$\mathrm{CC}+\mathrm{DESI\,DR2}+
\mathrm{PantheonPlus\&SH0ES}$.
The last column reports the minimum $\chi^2$.
}
	\label{tab:constraints_pantheonplus_shoes}
	
	\resizebox{\textwidth}{!}{%
		\begin{tabular}{lcccccccc}
			\toprule
			Model
			& $\Omega_m$
			& $h$
			& $M_B$
			& $w_0$
			& $w_a$
			& $\tilde{\sigma}_0$
			& Exponent
			& $\chi^2_{\min}$
			\\
			\midrule
			
			$\Lambda$CDM
			& $0.31 \pm 0.0078$
			& $0.70 \pm 0.0046$
			& $-19.37 \pm 0.015$
			& --
			& --
			& --
			& --
			& $1501.90$
			\\
			
			CPL
			& $0.33 \pm 0.0092$
			& $0.71 \pm 0.0076$
			& $-19.32 \pm 0.022$
			& $-0.87 \pm 0.068$
			& $-0.99 \pm 0.36$
			& --
			& --
			& $1491.03$
			\\
			
			Constant diffusion
			& $0.35 \pm 0.012$
			& $0.72 \pm 0.0074$
			& $-19.28 \pm 0.023$
			& --
			& --
            & $\begin{array}{c}
            0.079^{+0.019}_{-0.019} \\
            0.387^{+0.009}_{-0.012}
            \end{array}$
			& --
			& $1480.50$
			\\
			
			$\phi$-power $(\delta)$
			& $0.36 \pm 0.013$
			& $0.72 \pm 0.0086$
			& $-19.26 \pm 0.026$
			& --
			& --
			& $0.19 \pm 0.10$
			& $\delta=1.1 \pm 2.9$
			& $1479.70$
			\\
			
			$n$-power $(k)$
			& $0.35 \pm 0.013$
			& $0.73 \pm 0.0089$
			& $-19.26 \pm 0.027$
			& --
			& --
			& $0.22 \pm 0.096$
			& $k=-0.8 \pm 3.1$
			& $1479.47$
			\\
			
			$\rho$-power $(\lambda)$
			& $0.35 \pm 0.014$
			& $0.73 \pm 0.0078$
			& $-19.27 \pm 0.025$
			& --
			& --
			& $0.25 \pm 0.084$
			& $\lambda=-2.0 \pm 1.1$
			& $1479.46$
			\\
			
			$H$-power $(h_\sigma)$
			& $0.35 \pm 0.013$
			& $0.73 \pm 0.0080$
			& $-19.26 \pm 0.025$
			& --
			& --
			& $0.16 \pm 0.053$
			& $h_\sigma=-2.9 \pm 1.4$
			& $1479.42$
			\\
			
			\bottomrule
		\end{tabular}%
	}
\end{table*}

	\begin{table*}[t]
		\centering
\caption{
Bayesian model comparison for
$\mathrm{CC}+\mathrm{DESI\,DR2}+
\mathrm{PantheonPlus\&SH0ES}$.
All quantities are reported relative to $\Lambda$CDM.
}
		\label{tab:evidence_pantheonplus_shoes}
		
		\setlength{\tabcolsep}{10pt}
		\renewcommand{\arraystretch}{1.12}
		
		\begin{tabular}{lcccc}
			\toprule
			Model
			& $\chi^2_{\min}$
			& $\Delta\chi^2$
			& $\ln\mathcal{Z}$
			& $\Delta\ln\mathcal{Z}$
			\\
			\midrule
			
			$\Lambda$CDM
			& $1501.90$
			& $0.00$
			& $-764.683 \pm 0.153$
			& $0.000$
			\\
			
			CPL
			& $1491.03$
			& $-10.87$
			& $-764.469 \pm 0.179$
			& $+0.214 \pm 0.236$
			\\
			
			Constant diffusion
			& $1480.50$
			& $-21.40$
			& $-756.076 \pm 0.162$
			& $+8.607 \pm 0.223$
			\\
			
			$\phi$-power $(\delta)$
			& $1479.70$
			& $-22.20$
			& $-755.678 \pm 0.162$
			& $+9.005 \pm 0.223$
			\\
			
			$n$-power $(k)$
			& $1479.47$
			& $-22.43$
			& $-754.970 \pm 0.159$
			& $+9.713 \pm 0.221$
			\\
			
			$\rho$-power $(\lambda)$
			& $1479.46$
			& $-22.44$
			& $-755.358 \pm 0.161$
			& $+9.325 \pm 0.222$
			\\
			
			$H$-power $(h_\sigma)$
			& $1479.42$
			& $-22.48$
			& $-755.081 \pm 0.161$
			& $+9.602 \pm 0.222$
			\\
			
			\bottomrule
		\end{tabular}
	\end{table*}


The nested-sampling evidence derived for the dataset combination including the local SH0ES calibration leads to a qualitatively different conclusion from that obtained without local calibration. The corresponding Bayesian model-comparison results are summarized in Table~\ref{tab:evidence_pantheonplus_shoes}.

For the $\mathrm{CC}+\mathrm{DESI\,DR2}+\mathrm{PantheonPlus\&SH0ES}$ combination, the entire diffusion model family yields substantial improvements in the maximum likelihood ($-22.48 \le \Delta\chi^2 \le -21.40$), which translate into exceptionally strong Bayesian evidence:
\begin{equation}\label{eq:lnZ_shoes_range}
8.61 \lesssim \Delta\ln\mathcal{Z} \lesssim 9.71.
\end{equation}
According to the Jeffreys scale, values of $\Delta\ln\mathcal{Z}>5$ correspond to decisive evidence in favor of the extended model over $\Lambda$CDM. Therefore, the preference for matter diffusion under the SH0ES calibration remains robust even after fully accounting for the larger parameter volume. In stark contrast, the CPL parametrization yields only $\Delta\ln\mathcal{Z}=0.21\pm0.24$, despite its improvement in $\chi^2_{\mathrm{min}}$. Consequently, CPL receives no meaningful Bayesian preference over $\Lambda$CDM, demonstrating that a conventional dynamical dark-energy equation of state does not accommodate the local distance-ladder calibration as effectively as the matter-diffusion framework.

Within the diffusion family, the Hubble-dependent model, $\tilde{\sigma}_{(H)}$, provides the lowest $\chi^2_{\mathrm{min}}$, whereas the scale-factor-dependent model, $\tilde{\sigma}_{(n)}$, yields the highest Bayesian evidence. However, their log-evidence values differ by only $\Delta\ln\mathcal{Z}\simeq0.11$, which is smaller than the numerical uncertainty of the nested-sampling calculation and therefore does not represent a statistically meaningful distinction. Since both models introduce the same number of additional parameters and adopt comparable prior ranges for their respective exponents, the marginally higher evidence of $\tilde{\sigma}_{(n)}$ most likely reflects a slightly more favorable balance between goodness of fit and effective posterior-to-prior volume, rather than a genuine preference driven by substantially tighter parameter constraints.

More broadly, the evidence differences across the entire diffusion family remain small compared to their overwhelming collective preference over $\Lambda$CDM. The maximum spread is only $\Delta\ln\mathcal{Z} \simeq 1.1$, which corresponds, at most, to weak evidence on the Jeffreys scale. This reinforces the conclusion that the non-linear extensions do not provide a statistically compelling advantage over the single-parameter constant-diffusion model. Since these minor gains do not justify the introduction of an additional exponent parameter, the Bayesian evidence primarily favors the presence of a non-vanishing matter-diffusion interaction itself rather than a specific time-dependent functional form for the diffusion coefficient. From the standpoint of model parsimony, the single-parameter constant-diffusion model consequently emerges as a particularly well-motivated description of the current late-time observations.

\section{Discussion and Conclusions}\label{sec:conclusion}

The results obtained in this work reveal two main trends, demonstrating that the observational preference for matter diffusion depends strongly on whether the local SH0ES calibration is included in the dataset combination.

When the analysis is restricted to the baseline combination without local distance-ladder calibration ($\mathrm{CC}+\mathrm{DESI\,DR2}+\mathrm{PantheonPlus}$), the diffusion models provide only modest reductions in the minimum $\chi^2$ relative to $\Lambda$CDM. These improvements are comparable to those obtained for the CPL model and are not supported by the Bayesian evidence. The slightly negative values of $\Delta\ln\mathcal{Z}$ indicate that the additional flexibility of the extended models is insufficient to compensate for the increased prior volume. In this case, $\Lambda$CDM remains statistically competitive with both conventional dynamical dark-energy and matter-diffusion models.

A qualitatively different picture emerges when the SH0ES calibration is included ($\mathrm{CC}+\mathrm{DESI\,DR2}+\mathrm{PantheonPlus\&SH0ES}$). Here, all diffusion models produce substantial improvements in the best-fit $\chi^2$ relative to both $\Lambda$CDM and CPL, which translate into decisive Bayesian evidence. This demonstrates that the matter-diffusion framework modifies the late-time expansion history in a way that is particularly effective at accommodating datasets incorporating a local distance-ladder calibration.

However, the non-linear diffusion extensions do not clearly outperform the single-parameter constant-diffusion baseline. Although the four power-law cases (scaling with the scale factor, matter density, scalar-field density, or Hubble rate) yield slightly lower $\chi^2$ values, these incremental gains are small compared to the cost of introducing an additional exponent parameter. The constant-diffusion model already accounts for virtually all of the improvement in the fit. Consequently, the most conservative interpretation of the current late-time data is that the dominant statistical preference is driven by the presence of a non-vanishing matter-diffusion interaction itself, rather than by a specific time-dependent functional form of the diffusion coefficient.

The posterior distributions provide further insight into this behavior. Incorporating the SH0ES calibration systematically shifts the preferred Hubble parameter toward higher values across all models. Within the diffusion framework, this shift is accompanied by coordinated adjustments in the preferred matter density and diffusion amplitude, demonstrating how these parameters jointly determine the late-time expansion history. The parameter correlations visible in the corner plots show that the diffusion amplitude is strongly degenerate with the background cosmological parameters. Consequently, different combinations of $\Omega_m$, $H_0$, and $\tilde{\sigma}_0$ can provide comparably good descriptions of the current late-time observations.

Interestingly, the posterior distributions for the constant-diffusion model exhibit bimodality across both dataset compilations, appearing regardless of whether the local SH0ES calibration is included. This behavior stems directly from the non-linear way matter diffusion alters the background expansion. Because the diffusion term enters the matter and scalar-field continuity equations with opposite signs, it acts as a late-time energy transfer mechanism between the two sectors. Consequently, late-time distance-redshift observables can accommodate two distinct parameter configurations (a weak-diffusion branch and a strong-diffusion branch) while intermediate values of the present-day amplitude $\tilde{\sigma}_0$ yield a less optimal $H(z)$ trajectory. When the local SH0ES prior is incorporated, this bimodal structure becomes slightly tighter, reflecting the stronger observational pressure placed on the late-time expansion rate.

The non-linear diffusion extensions introduce a secondary degeneracy between the present-day amplitude $\tilde{\sigma}_0$ and the power-law exponent. The broad, weakly constrained posterior distributions for $k$, $\lambda$, $\delta$, and $h_\sigma$ indicate that current background data are primarily sensitive to the integrated effect of energy exchange on $H(z)$, rather than to the detailed time dependence of the interaction strength. This parameter degeneracy is directly reflected in the model-selection results: while $\tilde{\sigma}_{(H)}$ achieves the lowest $\chi^2_{\mathrm{min}}$, $\tilde{\sigma}_{(n)}$ yields the highest Bayesian evidence under the SH0ES calibration, with the differences among all non-linear parametrizations remaining minor compared to their shared preference over $\Lambda$CDM.
In the non-linear power-law extensions, the bimodal feature (present in the constant-diffusion case) is largely smeared out because the additional exponent introduces functional freedom for the interaction to evolve independently of its present-day normalization. The two distinct branches previously mentioned become continuously connected through a degeneracy between $\tilde{\sigma}_0$ and the corresponding power-law index. For instance, in the matter-density-dependent model $\tilde{\sigma}_{(\rho)}$, a strong negative correlation between $\tilde{\sigma}_0$ and $\lambda$ allows a larger present-day diffusion amplitude to be compensated by a more negative $\lambda$, which rapidly suppresses the interaction strength at higher redshifts. The transition from a bimodal posterior in the constant-diffusion case thus reflects the additional parameter freedom to redistribute the energy exchange across cosmic history.

The significant improvement in the fit obtained when the SH0ES calibration is included demonstrates that matter diffusion modifies the expansion history in a direction that effectively accommodates the locally calibrated distance scale. By allowing energy transfer between the matter and scalar-field sectors, the framework provides additional flexibility in the late-time expansion history without resorting to standard equation-of-state parametrizations such as CPL. The fact that the matter-diffusion framework outperforms CPL when the local calibration is included suggests that the background dynamics generated by matter-energy exchange provide a more effective description of the current low-redshift observations.

Nevertheless, these results should not be interpreted as a complete resolution of the Hubble tension. Since this study is restricted to late-time background observables, it does not include constraints from the CMB, weak gravitational lensing, or redshift-space distortions. Any viable cosmological model must also remain consistent with the growth of cosmic structures and other perturbation-level observables. Furthermore, the Bayesian evidence remains sensitive to the choice of prior volume, particularly for the broad posterior distributions of the power-law exponents. Testing the robustness of the observed preference by including early-Universe datasets, perturbation-level observables, and alternative supernova calibrations will therefore be essential to assess whether matter diffusion provides a viable global cosmological framework. Such an investigation, however, lies beyond the scope of the present work.

\acknowledgments

The authors thank Mahdi Najafi and Enrico Specogna for help and discussions at the beginning of the projects.
EDV is supported by a Royal Society Dorothy Hodgkin Research Fellowship.
L.A.E.\ acknowledges support from T\"{U}B\.{I}TAK through a postdoctoral researcher fellowship associated with Grant No.~124N627.
This article is based upon work from the COST Action CA21136 - ``Addressing observational tensions in cosmology with systematics and fundamental physics (CosmoVerse)'', supported by COST - ``European Cooperation in Science and Technology''. 
    
\appendix
	
\bibliographystyle{unsrt}
\bibliography{references}

@article{AtacamaCosmologyTelescope:2025blo,
    author = "Louis, Thibaut and others",
    collaboration = "Atacama Cosmology Telescope",
    title = "{The Atacama Cosmology Telescope: DR6 power spectra, likelihoods and {\ensuremath{\Lambda}}CDM parameters}",
    eprint = "2503.14452",
    archivePrefix = "arXiv",
    primaryClass = "astro-ph.CO",
    reportNumber = "FERMILAB-PUB-25-0071-PPD",
    doi = "10.1088/1475-7516/2025/11/062",
    journal = "JCAP",
    volume = "11",
    pages = "062",
    year = "2025"
}

@article{Planck:2018vyg,
    author = "Aghanim, N. and others",
    collaboration = "Planck",
    title = "{Planck 2018 results. VI. Cosmological parameters}",
    eprint = "1807.06209",
    archivePrefix = "arXiv",
    primaryClass = "astro-ph.CO",
    doi = "10.1051/0004-6361/201833910",
    journal = "Astron. Astrophys.",
    volume = "641",
    pages = "A6",
    year = "2020",
    note = "[Erratum: Astron.Astrophys. 652, C4 (2021)]"
}

@article{SPT-3G:2025bzu,
    author = "Camphuis, E. and others",
    collaboration = "SPT-3G",
    title = "{SPT-3G D1: CMB temperature and polarization power spectra and cosmology from 2019 and 2020 observations of the SPT-3G main field}",
    eprint = "2506.20707",
    archivePrefix = "arXiv",
    primaryClass = "astro-ph.CO",
    reportNumber = "FERMILAB-PUB-25-0144-PPD",
    doi = "10.1103/7wt3-9v2y",
    journal = "Phys. Rev. D",
    volume = "113",
    number = "8",
    pages = "083504",
    year = "2026"
}

@article{eBOSS:2020fvk,
    author = "de Mattia, Arnaud and others",
    collaboration = "eBOSS",
    title = "{The Completed SDSS-IV extended Baryon Oscillation Spectroscopic Survey: measurement of the BAO and growth rate of structure of the emission line galaxy sample from the anisotropic power spectrum between redshift 0.6 and 1.1}",
    eprint = "2007.09008",
    archivePrefix = "arXiv",
    primaryClass = "astro-ph.CO",
    doi = "10.1093/mnras/staa3891",
    journal = "Mon. Not. Roy. Astron. Soc.",
    volume = "501",
    number = "4",
    pages = "5616--5645",
    year = "2021"
}

@article{Brout:2022vxf,
    author = "Brout, Dillon and others",
    title = "{The Pantheon+ Analysis: Cosmological Constraints}",
    eprint = "2202.04077",
    archivePrefix = "arXiv",
    primaryClass = "astro-ph.CO",
    doi = "10.3847/1538-4357/ac8e04",
    journal = "Astrophys. J.",
    volume = "938",
    number = "2",
    pages = "110",
    year = "2022"
}

@inproceedings{Weinberg:2000yb,
    author = "Weinberg, Steven",
    title = "{The Cosmological constant problems}",
    booktitle = "{4th International Symposium on Sources and Detection of Dark Matter in the Universe (DM 2000)}",
    eprint = "astro-ph/0005265",
    archivePrefix = "arXiv",
    reportNumber = "UTTG-07-00",
    doi = "10.1007/978-3-662-04587-9_2",
    pages = "18--26",
    month = "2",
    year = "2000"
}

@article{Zlatev:1998tr,
    author = "Zlatev, Ivaylo and Wang, Li-Min and Steinhardt, Paul J.",
    title = "{Quintessence, cosmic coincidence, and the cosmological constant}",
    eprint = "astro-ph/9807002",
    archivePrefix = "arXiv",
    doi = "10.1103/PhysRevLett.82.896",
    journal = "Phys. Rev. Lett.",
    volume = "82",
    pages = "896--899",
    year = "1999"
}

@article{Carroll:2000fy,
    author = "Carroll, Sean M.",
    title = "{The Cosmological constant}",
    eprint = "astro-ph/0004075",
    archivePrefix = "arXiv",
    reportNumber = "EFI-2000-13",
    doi = "10.12942/lrr-2001-1",
    journal = "Living Rev. Rel.",
    volume = "4",
    pages = "1",
    year = "2001"
}

@article{Peebles:2002gy,
    author = "Peebles, P. J. E. and Ratra, Bharat",
    editor = "Hsu, Jong-Ping and Fine, D.",
    title = "{The Cosmological Constant and Dark Energy}",
    eprint = "astro-ph/0207347",
    archivePrefix = "arXiv",
    reportNumber = "KSUPT-02-3",
    doi = "10.1103/RevModPhys.75.559",
    journal = "Rev. Mod. Phys.",
    volume = "75",
    pages = "559--606",
    year = "2003"
}

@article{Verde:2019ivm,
    author = "Verde, L. and Treu, T. and Riess, A. G.",
    title = "{Tensions between the Early and the Late Universe}",
    eprint = "1907.10625",
    archivePrefix = "arXiv",
    primaryClass = "astro-ph.CO",
    doi = "10.1038/s41550-019-0902-0",
    journal = "Nature Astron.",
    volume = "3",
    pages = "891",
    year = "2019"
}

@article{DiValentino:2020zio,
    author = "Di Valentino, Eleonora and others",
    title = "{Snowmass2021 - Letter of interest cosmology intertwined II: The hubble constant tension}",
    eprint = "2008.11284",
    archivePrefix = "arXiv",
    primaryClass = "astro-ph.CO",
    reportNumber = "FERMILAB-PUB-21-590-PPD",
    doi = "10.1016/j.astropartphys.2021.102605",
    journal = "Astropart. Phys.",
    volume = "131",
    pages = "102605",
    year = "2021"
}

@article{DiValentino:2021izs,
    author = "Di Valentino, Eleonora and Mena, Olga and Pan, Supriya and Visinelli, Luca and Yang, Weiqiang and Melchiorri, Alessandro and Mota, David F. and Riess, Adam G. and Silk, Joseph",
    title = "{In the realm of the Hubble tension{\textemdash}a review of solutions}",
    eprint = "2103.01183",
    archivePrefix = "arXiv",
    primaryClass = "astro-ph.CO",
    reportNumber = "IPPP/20/108",
    doi = "10.1088/1361-6382/ac086d",
    journal = "Class. Quant. Grav.",
    volume = "38",
    number = "15",
    pages = "153001",
    year = "2021"
}

@article{Perivolaropoulos:2021jda,
    author = "Perivolaropoulos, Leandros and Skara, Foteini",
    title = "{Challenges for {\ensuremath{\Lambda}}CDM: An update}",
    eprint = "2105.05208",
    archivePrefix = "arXiv",
    primaryClass = "astro-ph.CO",
    doi = "10.1016/j.newar.2022.101659",
    journal = "New Astron. Rev.",
    volume = "95",
    pages = "101659",
    year = "2022"
}

@article{Schoneberg:2021qvd,
    author = {Sch{\"o}neberg, Nils and Franco Abell{\'a}n, Guillermo and P{\'e}rez S{\'a}nchez, Andrea and Witte, Samuel J. and Poulin, Vivian and Lesgourgues, Julien},
    title = "{The H0 Olympics: A fair ranking of proposed models}",
    eprint = "2107.10291",
    archivePrefix = "arXiv",
    primaryClass = "astro-ph.CO",
    doi = "10.1016/j.physrep.2022.07.001",
    journal = "Phys. Rept.",
    volume = "984",
    pages = "1--55",
    year = "2022"
}

@article{Shah:2021onj,
    author = "Shah, Paul and Lemos, Pablo and Lahav, Ofer",
    title = "{A buyer{\textquoteright}s guide to the Hubble constant}",
    eprint = "2109.01161",
    archivePrefix = "arXiv",
    primaryClass = "astro-ph.CO",
    doi = "10.1007/s00159-021-00137-4",
    journal = "Astron. Astrophys. Rev.",
    volume = "29",
    number = "1",
    pages = "9",
    year = "2021"
}

@article{Abdalla:2022yfr,
    author = "Abdalla, Elcio and others",
    title = "{Cosmology intertwined: A review of the particle physics, astrophysics, and cosmology associated with the cosmological tensions and anomalies}",
    eprint = "2203.06142",
    archivePrefix = "arXiv",
    primaryClass = "astro-ph.CO",
    reportNumber = "FERMILAB-CONF-22-192-SCD",
    doi = "10.1016/j.jheap.2022.04.002",
    journal = "JHEAp",
    volume = "34",
    pages = "49--211",
    year = "2022"
}

@article{DiValentino:2022fjm,
    author = "Di Valentino, Eleonora",
    title = "{Challenges of the Standard Cosmological Model}",
    doi = "10.3390/universe8080399",
    journal = "Universe",
    volume = "8",
    number = "8",
    pages = "399",
    year = "2022"
}

@article{Kamionkowski:2022pkx,
    author = "Kamionkowski, Marc and Riess, Adam G.",
    title = "{The Hubble Tension and Early Dark Energy}",
    eprint = "2211.04492",
    archivePrefix = "arXiv",
    primaryClass = "astro-ph.CO",
    doi = "10.1146/annurev-nucl-111422-024107",
    journal = "Ann. Rev. Nucl. Part. Sci.",
    volume = "73",
    pages = "153--180",
    year = "2023"
}

@article{Giare:2023xoc,
    author = "Giar{\`e}, William",
    title = "{CMB Anomalies and the Hubble Tension}",
    eprint = "2305.16919",
    archivePrefix = "arXiv",
    primaryClass = "astro-ph.CO",
    doi = "10.1007/978-981-99-0177-7_36",
    month = "5",
    year = "2023"
}

@article{Hu:2023jqc,
    author = "Hu, Jian-Ping and Wang, Fa-Yin",
    title = "{Hubble Tension: The Evidence of New Physics}",
    eprint = "2302.05709",
    archivePrefix = "arXiv",
    primaryClass = "astro-ph.CO",
    doi = "10.3390/universe9020094",
    journal = "Universe",
    volume = "9",
    number = "2",
    pages = "94",
    year = "2023"
}

@article{Verde:2023lmm,
    author = {Verde, Licia and Sch{\"o}neberg, Nils and Gil-Mar{\'\i}n, H{\'e}ctor},
    title = "{A Tale of Many H0}",
    eprint = "2311.13305",
    archivePrefix = "arXiv",
    primaryClass = "astro-ph.CO",
    doi = "10.1146/annurev-astro-052622-033813",
    journal = "Ann. Rev. Astron. Astrophys.",
    volume = "62",
    number = "1",
    pages = "287--331",
    year = "2024"
}

@book{DiValentino:2024yew,
    editor = "Di Valentino, Eleonora and Brout, Dillon",
    title = "{The Hubble Constant Tension}",
    doi = "10.1007/978-981-99-0177-7",
    isbn = "978-981-99-0176-0, 978-981-99-0179-1, 978-981-99-0177-7",
    publisher = "Springer",
    series = "Springer Series in Astrophysics and Cosmology",
    year = "2024"
}

@article{CosmoVerseNetwork:2025alb,
    author = "Di Valentino, Eleonora and others",
    collaboration = "CosmoVerse Network",
    title = "{The CosmoVerse White Paper: Addressing observational tensions in cosmology with systematics and fundamental physics}",
    eprint = "2504.01669",
    archivePrefix = "arXiv",
    primaryClass = "astro-ph.CO",
    doi = "10.1016/j.dark.2025.101965",
    journal = "Phys. Dark Univ.",
    volume = "49",
    pages = "101965",
    year = "2025"
}

@article{Ong:2025cwv,
    author = "Ong, Dily Duan Yi and Handley, Will",
    title = "{unimpeded: A Public Grid of Nested Sampling Chains for Cosmological Model Comparison and Tension Analysis}",
    eprint = "2511.04661",
    archivePrefix = "arXiv",
    primaryClass = "astro-ph.CO",
    month = "11",
    year = "2025"
}

@article{Cai:2026swf,
    author = "Cai, Rong-Gen and Wang, Shao-Jiang",
    title = "{The Hubble Tension: A Decade Review}",
    eprint = "2606.20434",
    archivePrefix = "arXiv",
    primaryClass = "astro-ph.CO",
    doi = "10.1088/1674-4527/ae842f",
    journal = "Res. Astron. Astrophys.",
    volume = "26",
    number = "8",
    pages = "084011",
    year = "2026"
}

@article{Planck:2018nkj,
    author = "Aghanim, N. and others",
    collaboration = "Planck",
    title = "{Planck 2018 results. I. Overview and the cosmological legacy of Planck}",
    eprint = "1807.06205",
    archivePrefix = "arXiv",
    primaryClass = "astro-ph.CO",
    doi = "10.1051/0004-6361/201833880",
    journal = "Astron. Astrophys.",
    volume = "641",
    pages = "A1",
    year = "2020"
}

@article{ACT:2025fju,
    author = "Louis, Thibaut and others",
    collaboration = "Atacama Cosmology Telescope",
    title = "{The Atacama Cosmology Telescope: DR6 power spectra, likelihoods and {\ensuremath{\Lambda}}CDM parameters}",
    eprint = "2503.14452",
    archivePrefix = "arXiv",
    primaryClass = "astro-ph.CO",
    reportNumber = "FERMILAB-PUB-25-0071-PPD",
    doi = "10.1088/1475-7516/2025/11/062",
    journal = "JCAP",
    volume = "11",
    pages = "062",
    year = "2025"
}

@article{Freedman:2020dne,
    author = "Freedman, Wendy L. and Madore, Barry F. and Hoyt, Taylor and Jang, In Sung and Beaton, Rachael and Lee, Myung Gyoon and Monson, Andrew and Neeley, Jill and Rich, Jeffrey",
    title = "{Calibration of the Tip of the Red Giant Branch}",
    eprint = "2002.01550",
    archivePrefix = "arXiv",
    primaryClass = "astro-ph.GA",
    doi = "10.3847/1538-4357/ab7339",
    journal = "Astrophys. J.",
    volume = "891",
    number = "1",
    pages = "57",
    year = "2020"
}

@article{Birrer:2020tax,
    author = "Birrer, S. and others",
    title = "{TDCOSMO - IV. Hierarchical time-delay cosmography {\textendash} joint inference of the Hubble constant and galaxy density profiles}",
    eprint = "2007.02941",
    archivePrefix = "arXiv",
    primaryClass = "astro-ph.CO",
    doi = "10.1051/0004-6361/202038861",
    journal = "Astron. Astrophys.",
    volume = "643",
    pages = "A165",
    year = "2020"
}

@article{Riess:2021jrx,
    author = "Riess, Adam G. and others",
    title = "{A Comprehensive Measurement of the Local Value of the Hubble Constant with 1 km/s/Mpc Uncertainty from the Hubble Space Telescope and the SH0ES Team}",
    eprint = "2112.04510",
    archivePrefix = "arXiv",
    primaryClass = "astro-ph.CO",
    doi = "10.3847/2041-8213/ac5c5b",
    journal = "Astrophys. J. Lett.",
    volume = "934",
    number = "1",
    pages = "L7",
    year = "2022"
}

@article{Anderson:2023aga,
    author = "Anderson, Richard I. and Koblischke, Nolan W. and Eyer, Laurent",
    title = "{Small-amplitude Red Giants Elucidate the Nature of the Tip of the Red Giant Branch as a Standard Candle}",
    eprint = "2303.04790",
    archivePrefix = "arXiv",
    primaryClass = "astro-ph.CO",
    doi = "10.3847/2041-8213/ad284d",
    journal = "Astrophys. J. Lett.",
    volume = "963",
    number = "2",
    pages = "L43",
    year = "2024"
}

@article{Scolnic:2023mrv,
    author = "Scolnic, D. and Riess, A. G. and Wu, J. and Li, S. and Anand, G. S. and Beaton, R. and Casertano, S. and Anderson, R. I. and Dhawan, S. and Ke, X.",
    title = "{CATS: The Hubble Constant from Standardized TRGB and Type Ia Supernova Measurements}",
    eprint = "2304.06693",
    archivePrefix = "arXiv",
    primaryClass = "astro-ph.CO",
    doi = "10.3847/2041-8213/ace978",
    journal = "Astrophys. J. Lett.",
    volume = "954",
    number = "1",
    pages = "L31",
    year = "2023"
}

@article{Jones:2022mvo,
    author = "Jones, D. O. and others",
    title = "{Cosmological Results from the RAISIN Survey: Using Type Ia Supernovae in the Near Infrared as a Novel Path to Measure the Dark Energy Equation of State}",
    eprint = "2201.07801",
    archivePrefix = "arXiv",
    primaryClass = "astro-ph.CO",
    doi = "10.3847/1538-4357/ac755b",
    journal = "Astrophys. J.",
    volume = "933",
    number = "2",
    pages = "172",
    year = "2022"
}

@article{Anand:2021sum,
    author = "Anand, Gagandeep S. and Tully, R. Brent and Rizzi, Luca and Riess, Adam G. and Yuan, Wenlong",
    title = "{Comparing Tip of the Red Giant Branch Distance Scales: An Independent Reduction of the Carnegie-Chicago Hubble Program and the Value of the Hubble Constant}",
    eprint = "2108.00007",
    archivePrefix = "arXiv",
    primaryClass = "astro-ph.CO",
    doi = "10.3847/1538-4357/ac68df",
    journal = "Astrophys. J.",
    volume = "932",
    number = "1",
    pages = "15",
    year = "2022"
}

@article{Freedman:2021ahq,
    author = "Freedman, Wendy L.",
    title = "{Measurements of the Hubble Constant: Tensions in Perspective}",
    eprint = "2106.15656",
    archivePrefix = "arXiv",
    primaryClass = "astro-ph.CO",
    doi = "10.3847/1538-4357/ac0e95",
    journal = "Astrophys. J.",
    volume = "919",
    number = "1",
    pages = "16",
    year = "2021"
}

@article{Uddin:2023iob,
    author = "Uddin, Syed A. and others",
    title = "{Carnegie Supernova Project I and II: Measurements of H $_{0}$ Using Cepheid, Tip of the Red Giant Branch, and Surface Brightness Fluctuation Distance Calibration to Type Ia Supernovae*}",
    eprint = "2308.01875",
    archivePrefix = "arXiv",
    primaryClass = "astro-ph.CO",
    doi = "10.3847/1538-4357/ad3e63",
    journal = "Astrophys. J.",
    volume = "970",
    number = "1",
    pages = "72",
    year = "2024"
}

@article{Huang:2023frr,
    author = "Huang, Caroline D. and others",
    title = "{The Mira Distance to M101 and a 4{\%} Measurement of H $_{0}$}",
    eprint = "2312.08423",
    archivePrefix = "arXiv",
    primaryClass = "astro-ph.CO",
    doi = "10.3847/1538-4357/ad1ff8",
    journal = "Astrophys. J.",
    volume = "963",
    number = "2",
    pages = "83",
    year = "2024"
}

@article{Li:2024yoe,
    author = "Li, Siyang and Riess, Adam G. and Casertano, Stefano and Anand, Gagandeep S. and Scolnic, Daniel M. and Yuan, Wenlong and Breuval, Louise and Huang, Caroline D.",
    title = "{Reconnaissance with JWST of the J-region Asymptotic Giant Branch in Distance Ladder Galaxies: From Irregular Luminosity Functions to Approximation of the Hubble Constant}",
    eprint = "2401.04777",
    archivePrefix = "arXiv",
    primaryClass = "astro-ph.CO",
    doi = "10.3847/1538-4357/ad2f2b",
    journal = "Astrophys. J.",
    volume = "966",
    number = "1",
    pages = "20",
    year = "2024"
}

@article{Pesce:2020xfe,
    author = "Pesce, D. W. and others",
    title = "{The Megamaser Cosmology Project. XIII. Combined Hubble constant constraints}",
    eprint = "2001.09213",
    archivePrefix = "arXiv",
    primaryClass = "astro-ph.CO",
    doi = "10.3847/2041-8213/ab75f0",
    journal = "Astrophys. J. Lett.",
    volume = "891",
    number = "1",
    pages = "L1",
    year = "2020"
}

@article{Kourkchi:2020iyz,
    author = "Kourkchi, Ehsan and Tully, R. Brent and Anand, Gagandeep S. and Courtois, Helene M. and Dupuy, Alexandra and Neill, James D. and Rizzi, Luca and Seibert, Mark",
    title = "{Cosmicflows-4: The Calibration of Optical and Infrared Tully{\textendash}Fisher Relations}",
    eprint = "2004.14499",
    archivePrefix = "arXiv",
    primaryClass = "astro-ph.GA",
    doi = "10.3847/1538-4357/ab901c",
    journal = "Astrophys. J.",
    volume = "896",
    number = "1",
    pages = "3",
    year = "2020"
}

@article{Schombert:2020pxm,
    author = "Schombert, James and McGaugh, Stacy and Lelli, Federico",
    title = "{Using the Baryonic Tully{\textendash}Fisher Relation to Measure H o}",
    eprint = "2006.08615",
    archivePrefix = "arXiv",
    primaryClass = "astro-ph.CO",
    doi = "10.3847/1538-3881/ab9d88",
    journal = "Astron. J.",
    volume = "160",
    number = "2",
    pages = "71",
    year = "2020"
}

@article{Blakeslee:2021rqi,
    author = "Blakeslee, John P. and Jensen, Joseph B. and Ma, Chung-Pei and Milne, Peter A. and Greene, Jenny E.",
    title = "{The Hubble Constant from Infrared Surface Brightness Fluctuation Distances}",
    eprint = "2101.02221",
    archivePrefix = "arXiv",
    primaryClass = "astro-ph.CO",
    doi = "10.3847/1538-4357/abe86a",
    journal = "Astrophys. J.",
    volume = "911",
    number = "1",
    pages = "65",
    year = "2021"
}

@article{deJaeger:2022lit,
    author = "de Jaeger, T. and Galbany, L. and Riess, A. G. and Stahl, B. E. and Shappee, B. J. and Filippenko, A. V. and Zheng, W.",
    title = "{A 5~per{\,}cent measurement of the Hubble{\textendash}Lema{\^\i}tre constant from Type II supernovae}",
    eprint = "2203.08974",
    archivePrefix = "arXiv",
    primaryClass = "astro-ph.CO",
    doi = "10.1093/mnras/stac1661",
    journal = "Mon. Not. Roy. Astron. Soc.",
    volume = "514",
    number = "3",
    pages = "4620--4628",
    year = "2022"
}

@article{Murakami:2023xuy,
    author = "Murakami, Yukei S. and Riess, Adam G. and Stahl, Benjamin E. and Kenworthy, W. D'Arcy and Pluck, Dahne-More A. and Macoretta, Antonella and Brout, Dillon and Jones, David O. and Scolnic, Dan M. and Filippenko, Alexei V.",
    title = "{Leveraging SN Ia spectroscopic similarity to improve the measurement of H $_{0}$}",
    eprint = "2306.00070",
    archivePrefix = "arXiv",
    primaryClass = "astro-ph.CO",
    doi = "10.1088/1475-7516/2023/11/046",
    journal = "JCAP",
    volume = "11",
    pages = "046",
    year = "2023"
}

@article{Breuval:2024lsv,
    author = "Breuval, Louise and Riess, Adam G. and Casertano, Stefano and Yuan, Wenlong and Macri, Lucas M. and Romaniello, Martino and Murakami, Yukei S. and Scolnic, Daniel and Anand, Gagandeep S. and Soszy{\'n}ski, Igor",
    title = "{Small Magellanic Cloud Cepheids Observed with the Hubble Space Telescope Provide a New Anchor for the SH0ES Distance Ladder}",
    eprint = "2404.08038",
    archivePrefix = "arXiv",
    primaryClass = "astro-ph.CO",
    doi = "10.3847/1538-4357/ad630e",
    journal = "Astrophys. J.",
    volume = "973",
    number = "1",
    pages = "30",
    year = "2024"
}

@article{Freedman:2024eph,
    author = "Freedman, Wendy L. and Madore, Barry F. and Hoyt, Taylor J. and Jang, In Sung and Lee, Abigail J. and Owens, Kayla A.",
    title = "{Status Report on the Chicago-Carnegie Hubble Program (CCHP): Measurement of the Hubble Constant Using the Hubble and James Webb Space Telescopes}",
    eprint = "2408.06153",
    archivePrefix = "arXiv",
    primaryClass = "astro-ph.CO",
    doi = "10.3847/1538-4357/adce78",
    journal = "Astrophys. J.",
    volume = "985",
    number = "2",
    pages = "203",
    year = "2025",
    note = "[Erratum: Astrophys.J. 993, 252 (2025)]"
}

@article{Riess:2024vfa,
    author = "Riess, Adam G. and others",
    title = "{JWST Validates HST Distance Measurements: Selection of Supernova Subsample Explains Differences in JWST Estimates of Local H $_{0}$}",
    eprint = "2408.11770",
    archivePrefix = "arXiv",
    primaryClass = "astro-ph.CO",
    doi = "10.3847/1538-4357/ad8c21",
    journal = "Astrophys. J.",
    volume = "977",
    number = "1",
    pages = "120",
    year = "2024"
}

@article{Vogl:2024bum,
    author = "Vogl, Christian and others",
    title = "{No rungs attached: A distance-ladder-free determination of the Hubble constant through type II supernova spectral modelling}",
    eprint = "2411.04968",
    archivePrefix = "arXiv",
    primaryClass = "astro-ph.CO",
    doi = "10.1051/0004-6361/202452910",
    journal = "Astron. Astrophys.",
    volume = "702",
    pages = "A41",
    year = "2025"
}

@article{Scolnic:2024hbh,
    author = "Scolnic, Daniel and others",
    title = "{The Hubble Tension in Our Own Backyard: DESI and the Nearness of the Coma Cluster}",
    eprint = "2409.14546",
    archivePrefix = "arXiv",
    primaryClass = "astro-ph.CO",
    doi = "10.3847/2041-8213/ada0bd",
    journal = "Astrophys. J. Lett.",
    volume = "979",
    number = "1",
    pages = "L9",
    year = "2025"
}

@article{Said:2024pwm,
    author = "Said, Khaled and others",
    title = "{DESI peculiar velocity survey {\textendash} Fundamental Plane}",
    eprint = "2408.13842",
    archivePrefix = "arXiv",
    primaryClass = "astro-ph.CO",
    doi = "10.1093/mnras/staf700",
    journal = "Mon. Not. Roy. Astron. Soc.",
    volume = "539",
    number = "4",
    pages = "3627--3644",
    year = "2025"
}

@article{Boubel:2024cqw,
    author = "Boubel, Paula and Colless, Matthew and Said, Khaled and Staveley-Smith, Lister",
    title = "{An improved Tully{\textendash}Fisher estimate of H0}",
    eprint = "2408.03660",
    archivePrefix = "arXiv",
    primaryClass = "astro-ph.CO",
    doi = "10.1093/mnras/stae1925",
    journal = "Mon. Not. Roy. Astron. Soc.",
    volume = "533",
    number = "2",
    pages = "1550--1559",
    year = "2024"
}

@article{Scolnic:2024oth,
    author = "Scolnic, Daniel and Boubel, Paula and Byrne, Jakob and Riess, Adam G. and Anand, Gagandeep S.",
    title = "{Calibrating the Tully-Fisher Relation to Measure the Hubble Constant}",
    eprint = "2412.08449",
    archivePrefix = "arXiv",
    primaryClass = "astro-ph.CO",
    month = "12",
    year = "2024"
}

@article{Li:2025ife,
    author = "Li, Siyang and Riess, Adam G. and Scolnic, Daniel and Casertano, Stefano and Anand, Gagandeep S.",
    title = "{JAGB 2.0: Improved Constraints on the J-region Asymptotic Giant Branch{\textendash}based Hubble Constant from an Expanded Sample of JWST Observations}",
    eprint = "2502.05259",
    archivePrefix = "arXiv",
    primaryClass = "astro-ph.CO",
    doi = "10.3847/1538-4357/addd0c",
    journal = "Astrophys. J.",
    volume = "988",
    number = "1",
    pages = "97",
    year = "2025"
}

@article{Jensen:2025aai,
    author = "Jensen, Joseph B. and Blakeslee, John P. and Cantiello, Michele and Cowles, Mikaela and Anand, Gagandeep S. and Tully, R. Brent and Kourkchi, Ehsan and Raimondo, Gabriella",
    title = "{The TRGB-SBF Project. III. Refining the HST Surface Brightness Fluctuation Distance Scale Calibration with JWST}",
    eprint = "2502.15935",
    archivePrefix = "arXiv",
    primaryClass = "astro-ph.CO",
    doi = "10.3847/1538-4357/addfd6",
    month = "6",
    year = "2025"
}

@article{Riess:2025chq,
    author = "Riess, Adam G. and others",
    title = "{The Perfect Host: JWST Cepheid Observations in a Background-free Type Ia Supernova Host Confirm No Bias in Hubble-constant Measurements}",
    eprint = "2509.01667",
    archivePrefix = "arXiv",
    primaryClass = "astro-ph.CO",
    doi = "10.3847/2041-8213/ae0ad6",
    journal = "Astrophys. J. Lett.",
    volume = "992",
    number = "2",
    pages = "L34",
    year = "2025"
}

@article{Benisty:2025tct,
    author = "Benisty, David and Wagner, Jenny and Haridasu, Sandeep and Salucci, Paolo",
    title = "{Unveiling the Coma cluster structure: from the core to the Hubble flow}",
    eprint = "2504.04135",
    archivePrefix = "arXiv",
    primaryClass = "astro-ph.CO",
    doi = "10.1088/1475-7516/2026/06/005",
    journal = "JCAP",
    volume = "06",
    pages = "005",
    year = "2026"
}

@article{Newman:2025gwg,
    author = "Newman, Max J. B. and others",
    title = "{Tip of the Red Giant Branch Distances to NGC 1316, NGC 1380, NGC 1404, and NGC 4457: A Pilot Study of a Parallel Distance Ladder Using Type Ia Supernovae in Early-type Host Galaxies}",
    eprint = "2508.20023",
    archivePrefix = "arXiv",
    primaryClass = "astro-ph.CO",
    doi = "10.3847/1538-4357/ae8090",
    journal = "Astrophys. J.",
    volume = "1006",
    number = "2",
    pages = "178",
    year = "2026"
}

@article{Stiskalek:2025ktq,
    author = "Stiskalek, Richard and Desmond, Harry and Tsaprazi, Eleni and Heavens, Alan and Lavaux, Guilhem and McAlpine, Stuart and Jasche, Jens",
    title = "{1.8~per{\,}cent measurement of H0 from Cepheids alone}",
    eprint = "2509.09665",
    archivePrefix = "arXiv",
    primaryClass = "astro-ph.CO",
    doi = "10.1093/mnras/staf2260",
    journal = "Mon. Not. Roy. Astron. Soc.",
    volume = "546",
    number = "2",
    pages = "staf2260",
    year = "2026"
}

@article{Agrawal:2025tuv,
    author = "Agrawal, Aadya and others",
    title = "{Testing Lens Models of PLCK G165.7+67.0 Using Lensed Supernova H0pe}",
    eprint = "2510.07637",
    archivePrefix = "arXiv",
    primaryClass = "astro-ph.CO",
    doi = "10.3847/1538-4357/ae5f65",
    journal = "Astrophys. J.",
    volume = "1002",
    number = "2",
    pages = "187",
    year = "2026"
}

@article{Bhardwaj:2025kbw,
    author = "Bhardwaj, Anupam and Matsunaga, Noriyuki and Huang, Caroline D. and Riess, Adam G. and Rejkuba, Marina",
    title = "{Absolute Calibration of Cluster Mira Variables to Provide a New Anchor for the Hubble Constant Determination}",
    eprint = "2507.10658",
    archivePrefix = "arXiv",
    primaryClass = "astro-ph.GA",
    doi = "10.3847/1538-4357/adf20b",
    journal = "Astrophys. J.",
    volume = "990",
    number = "1",
    pages = "63",
    year = "2025"
}

@article{H0DN:2025lyy,
    author = "Casertano, Stefano and others",
    collaboration = "H0DN",
    title = "{The Local Distance Network: A community consensus report on the measurement of the Hubble constant at {\ensuremath{\sim}}1{\%} precision}",
    eprint = "2510.23823",
    archivePrefix = "arXiv",
    primaryClass = "astro-ph.CO",
    doi = "10.1051/0004-6361/202557993",
    journal = "Astron. Astrophys.",
    volume = "708",
    pages = "A166",
    year = "2026"
}

@article{DESI:2025fii,
    author = "Lodha, K. and others",
    collaboration = "DESI",
    title = "{Extended dark energy analysis using DESI DR2 BAO measurements}",
    eprint = "2503.14743",
    archivePrefix = "arXiv",
    primaryClass = "astro-ph.CO",
    reportNumber = "FERMILAB-PUB-25-0164-PPD",
    doi = "10.1103/w4c6-1r5j",
    journal = "Phys. Rev. D",
    volume = "112",
    number = "8",
    pages = "083511",
    year = "2025"
}

@article{DESI:2024mwx,
    author = "Adame, A. G. and others",
    collaboration = "DESI",
    title = "{DESI 2024 VI: cosmological constraints from the measurements of baryon acoustic oscillations}",
    eprint = "2404.03002",
    archivePrefix = "arXiv",
    primaryClass = "astro-ph.CO",
    reportNumber = "FERMILAB-PUB-24-0154-PPD",
    doi = "10.1088/1475-7516/2025/02/021",
    journal = "JCAP",
    volume = "02",
    pages = "021",
    year = "2025"
}

@article{DESI:2025zgx,
    author = "Abdul Karim, M. and others",
    collaboration = "DESI",
    title = "{DESI DR2 results. II. Measurements of baryon acoustic oscillations and cosmological constraints}",
    eprint = "2503.14738",
    archivePrefix = "arXiv",
    primaryClass = "astro-ph.CO",
    reportNumber = "FERMILAB-PUB-25-0169-PPD",
    doi = "10.1103/tr6y-kpc6",
    journal = "Phys. Rev. D",
    volume = "112",
    number = "8",
    pages = "083515",
    year = "2025"
}

@article{DESI:2024kob,
    author = "Lodha, K. and others",
    collaboration = "DESI",
    title = "{DESI 2024: Constraints on physics-focused aspects of dark energy using DESI DR1 BAO data}",
    eprint = "2405.13588",
    archivePrefix = "arXiv",
    primaryClass = "astro-ph.CO",
    reportNumber = "FERMILAB-PUB-24-0756-PPD",
    doi = "10.1103/PhysRevD.111.023532",
    journal = "Phys. Rev. D",
    volume = "111",
    number = "2",
    pages = "023532",
    year = "2025"
}

@article{Hoyt:2026fve,
    author = "Hoyt, Taylor J. and Rubin, David and Aldering, Greg and Perlmutter, Saul and Cuceu, Andrei and Gupta, Ravi",
    title = "{Union3.1: Self-consistent Measurements of Host Galaxy Properties for 2000 Type Ia Supernovae}",
    eprint = "2601.19424",
    archivePrefix = "arXiv",
    primaryClass = "astro-ph.CO",
    month = "1",
    year = "2026"
}

@article{DES:2025sig,
    author = "Popovic, B. and others",
    collaboration = "DES",
    title = "{The Dark Energy Survey supernova program: a reanalysis of cosmology results and evidence for evolving dark energy with an updated Type Ia supernova calibration}",
    eprint = "2511.07517",
    archivePrefix = "arXiv",
    primaryClass = "astro-ph.CO",
    reportNumber = "FERMILAB-PUB-25-0842-CSAID-PPD",
    doi = "10.1093/mnras/stag632",
    journal = "Mon. Not. Roy. Astron. Soc.",
    volume = "548",
    number = "4",
    pages = "stag632",
    year = "2026"
}

@article{Scolnic:2021amr,
    author = "Scolnic, Dan and others",
    title = "{The Pantheon+ Analysis: The Full Data Set and Light-curve Release}",
    eprint = "2112.03863",
    archivePrefix = "arXiv",
    primaryClass = "astro-ph.CO",
    doi = "10.3847/1538-4357/ac8b7a",
    journal = "Astrophys. J.",
    volume = "938",
    number = "2",
    pages = "113",
    year = "2022"
}

@article{Cortes:2024lgw,
    author = "Cort{\^e}s, Marina and Liddle, Andrew R.",
    title = "{Interpreting DESI's evidence for evolving dark energy}",
    eprint = "2404.08056",
    archivePrefix = "arXiv",
    primaryClass = "astro-ph.CO",
    doi = "10.1088/1475-7516/2024/12/007",
    journal = "JCAP",
    volume = "12",
    pages = "007",
    year = "2024"
}

@article{Shlivko:2024llw,
    author = "Shlivko, David and Steinhardt, Paul J.",
    title = "{Assessing observational constraints on dark energy}",
    eprint = "2405.03933",
    archivePrefix = "arXiv",
    primaryClass = "astro-ph.CO",
    doi = "10.1016/j.physletb.2024.138826",
    journal = "Phys. Lett. B",
    volume = "855",
    pages = "138826",
    year = "2024"
}

@article{Luongo:2024fww,
    author = "Luongo, Orlando and Muccino, Marco",
    title = "{Model-independent cosmographic constraints from DESI 2024}",
    eprint = "2404.07070",
    archivePrefix = "arXiv",
    primaryClass = "astro-ph.CO",
    doi = "10.1051/0004-6361/202450512",
    journal = "Astron. Astrophys.",
    volume = "690",
    pages = "A40",
    year = "2024"
}

@article{Gialamas:2024lyw,
    author = {Gialamas, Ioannis D. and H{\"u}tsi, Gert and Kannike, Kristjan and Racioppi, Antonio and Raidal, Martti and Vasar, Martin and Veerm{\"a}e, Hardi},
    title = "{Interpreting DESI 2024 BAO: Late-time dynamical dark energy or a local effect?}",
    eprint = "2406.07533",
    archivePrefix = "arXiv",
    primaryClass = "astro-ph.CO",
    doi = "10.1103/PhysRevD.111.043540",
    journal = "Phys. Rev. D",
    volume = "111",
    number = "4",
    pages = "043540",
    year = "2025"
}

@article{Wang:2024dka,
    author = "Wang, Hao and Piao, Yun-Song",
    title = "{Dark energy in light of DESI DR1 and Hubble tension}",
    eprint = "2404.18579",
    archivePrefix = "arXiv",
    primaryClass = "astro-ph.CO",
    doi = "10.1016/j.physletb.2026.140180",
    journal = "Phys. Lett. B",
    volume = "873",
    pages = "140180",
    year = "2026"
}

@article{Ye:2024ywg,
    author = "Ye, Gen and Martinelli, Matteo and Hu, Bin and Silvestri, Alessandra",
    title = "{Hints of Nonminimally Coupled Gravity in DESI 2024 Baryon Acoustic Oscillation Measurements}",
    eprint = "2407.15832",
    archivePrefix = "arXiv",
    primaryClass = "astro-ph.CO",
    doi = "10.1103/PhysRevLett.134.181002",
    journal = "Phys. Rev. Lett.",
    volume = "134",
    number = "18",
    pages = "181002",
    year = "2025"
}

@article{Tada:2024znt,
    author = "Tada, Yuichiro and Terada, Takahiro",
    title = "{Quintessential interpretation of the evolving dark energy in light of DESI observations}",
    eprint = "2404.05722",
    archivePrefix = "arXiv",
    primaryClass = "astro-ph.CO",
    doi = "10.1103/PhysRevD.109.L121305",
    journal = "Phys. Rev. D",
    volume = "109",
    number = "12",
    pages = "L121305",
    year = "2024"
}

@article{Carloni:2024zpl,
    author = "Carloni, Youri and Luongo, Orlando and Muccino, Marco",
    title = "{Does dark energy really revive using DESI 2024 data?}",
    eprint = "2404.12068",
    archivePrefix = "arXiv",
    primaryClass = "astro-ph.CO",
    doi = "10.1103/PhysRevD.111.023512",
    journal = "Phys. Rev. D",
    volume = "111",
    number = "2",
    pages = "023512",
    year = "2025"
}

@article{Chan-GyungPark:2024mlx,
    author = "Park, Chan-Gyung and de Cruz P{\'e}rez, Javier and Ratra, Bharat",
    title = "{Using non-DESI data to confirm and strengthen the DESI 2024 spatially flat w0waCDM cosmological parametrization result}",
    eprint = "2405.00502",
    archivePrefix = "arXiv",
    primaryClass = "astro-ph.CO",
    doi = "10.1103/PhysRevD.110.123533",
    journal = "Phys. Rev. D",
    volume = "110",
    number = "12",
    pages = "123533",
    year = "2024"
}

@article{Bhattacharya:2024hep,
    author = "Bhattacharya, Sukannya and Borghetto, Giulia and Malhotra, Ameek and Parameswaran, Susha and Tasinato, Gianmassimo and Zavala, Ivonne",
    title = "{Cosmological constraints on curved quintessence}",
    eprint = "2405.17396",
    archivePrefix = "arXiv",
    primaryClass = "astro-ph.CO",
    doi = "10.1088/1475-7516/2024/09/073",
    journal = "JCAP",
    volume = "09",
    pages = "073",
    year = "2024"
}

@article{Reboucas:2024smm,
    author = "Rebou{\c{c}}as, Jo{\~a}o and de Souza, Diogo H. F. and Zhong, Kunhao and Miranda, Vivian and Rosenfeld, Rogerio",
    title = "{Investigating late-time dark energy and massive neutrinos in light of DESI Y1 BAO}",
    eprint = "2408.14628",
    archivePrefix = "arXiv",
    primaryClass = "astro-ph.CO",
    doi = "10.1088/1475-7516/2025/02/024",
    journal = "JCAP",
    volume = "02",
    pages = "024",
    year = "2025"
}

@article{Najafi:2024qzm,
    author = "Najafi, Mahdi and Pan, Supriya and Di Valentino, Eleonora and Firouzjaee, Javad T.",
    title = "{Dynamical dark energy confronted with multiple CMB missions}",
    eprint = "2407.14939",
    archivePrefix = "arXiv",
    primaryClass = "astro-ph.CO",
    doi = "10.1016/j.dark.2024.101539",
    journal = "Phys. Dark Univ.",
    volume = "45",
    pages = "101539",
    year = "2024"
}

@article{Giare:2024gpk,
    author = "Giar{\`e}, William and Najafi, Mahdi and Pan, Supriya and Di Valentino, Eleonora and Firouzjaee, Javad T.",
    title = "{Robust preference for Dynamical Dark Energy in DESI BAO and SN measurements}",
    eprint = "2407.16689",
    archivePrefix = "arXiv",
    primaryClass = "astro-ph.CO",
    doi = "10.1088/1475-7516/2024/10/035",
    journal = "JCAP",
    volume = "10",
    pages = "035",
    year = "2024"
}

@article{Giare:2024ocw,
    author = "Giar{\`e}, William",
    title = "{Dynamical dark energy beyond Planck? Constraints from multiple CMB probes, DESI BAO, and type-Ia supernovae}",
    eprint = "2409.17074",
    archivePrefix = "arXiv",
    primaryClass = "astro-ph.CO",
    doi = "10.1103/ss37-cxhn",
    journal = "Phys. Rev. D",
    volume = "112",
    number = "2",
    pages = "023508",
    year = "2025"
}

@article{Jiang:2024xnu,
    author = "Jiang, Jun-Qian and Pedrotti, Davide and da Costa, Simony Santos and Vagnozzi, Sunny",
    title = "{Nonparametric late-time expansion history reconstruction and implications for the Hubble tension in light of recent DESI and type Ia supernovae data}",
    eprint = "2408.02365",
    archivePrefix = "arXiv",
    primaryClass = "astro-ph.CO",
    doi = "10.1103/PhysRevD.110.123519",
    journal = "Phys. Rev. D",
    volume = "110",
    number = "12",
    pages = "123519",
    year = "2024"
}

@article{RoyChoudhury:2024wri,
    author = "Roy Choudhury, Shouvik and Okumura, Teppei",
    title = "{Updated Cosmological Constraints in Extended Parameter Space with Planck PR4, DESI Baryon Acoustic Oscillations, and Supernovae: Dynamical Dark Energy, Neutrino Masses, Lensing Anomaly, and the Hubble Tension}",
    eprint = "2409.13022",
    archivePrefix = "arXiv",
    primaryClass = "astro-ph.CO",
    doi = "10.3847/2041-8213/ad8c26",
    journal = "Astrophys. J. Lett.",
    volume = "976",
    number = "1",
    pages = "L11",
    year = "2024"
}

@article{Giare:2024oil,
    author = "Giar{\`e}, William",
    title = "{Dynamical dark energy beyond Planck? Constraints from multiple CMB probes, DESI BAO, and type-Ia supernovae}",
    eprint = "2409.17074",
    archivePrefix = "arXiv",
    primaryClass = "astro-ph.CO",
    doi = "10.1103/ss37-cxhn",
    journal = "Phys. Rev. D",
    volume = "112",
    number = "2",
    pages = "023508",
    year = "2025"
}

@article{Giare:2025pzu,
    author = "Giar{\`e}, William and Mahassen, Tariq and Di Valentino, Eleonora and Pan, Supriya",
    title = "{An overview of what current data can (and cannot yet) say about evolving dark energy}",
    eprint = "2502.10264",
    archivePrefix = "arXiv",
    primaryClass = "astro-ph.CO",
    doi = "10.1016/j.dark.2025.101906",
    journal = "Phys. Dark Univ.",
    volume = "48",
    pages = "101906",
    year = "2025"
}

@article{Kessler:2025kju,
    author = "Kessler, Daniel A. and Escamilla, Luis A. and Pan, Supriya and Di Valentino, Eleonora",
    title = "{One-parameter dynamical dark energy: Hints for oscillations}",
    eprint = "2504.00776",
    archivePrefix = "arXiv",
    primaryClass = "astro-ph.CO",
    month = "4",
    year = "2025"
}

@article{RoyChoudhury:2025dhe,
    author = "Roy Choudhury, Shouvik",
    title = "{Cosmology in Extended Parameter Space with DESI Data Release 2 Baryon Acoustic Oscillations: A 2{\ensuremath{\sigma}}+ Detection of Nonzero Neutrino Masses with an Update on Dynamical Dark Energy and Lensing Anomaly}",
    eprint = "2504.15340",
    archivePrefix = "arXiv",
    primaryClass = "astro-ph.CO",
    doi = "10.3847/2041-8213/ade1cc",
    journal = "Astrophys. J. Lett.",
    volume = "986",
    number = "2",
    pages = "L31",
    year = "2025",
    note = "[Erratum: Astrophys.J.Lett. 1001, L25 (2026), Erratum: Astrophys.J. 1001, L25 (2026)]"
}

@article{Scherer:2025esj,
    author = "Scherer, Mateus and Sabogal, Miguel A. and Nunes, Rafael C. and De Felice, Antonio",
    title = "{Challenging the {\ensuremath{\Lambda}}CDM model: 5{\ensuremath{\sigma}} evidence for a dynamical dark energy late-time transition}",
    eprint = "2504.20664",
    archivePrefix = "arXiv",
    primaryClass = "astro-ph.CO",
    doi = "10.1103/n86r-sjgm",
    journal = "Phys. Rev. D",
    volume = "112",
    number = "4",
    pages = "043513",
    year = "2025"
}

@article{Wolf:2025jlc,
    author = "Wolf, William J. and Garc{\'\i}a-Garc{\'\i}a, Carlos and Ferreira, Pedro G.",
    title = "{Robustness of dark energy phenomenology across different parameterizations}",
    eprint = "2502.04929",
    archivePrefix = "arXiv",
    primaryClass = "astro-ph.CO",
    doi = "10.1088/1475-7516/2025/05/034",
    journal = "JCAP",
    volume = "05",
    pages = "034",
    year = "2025"
}

@article{Santos:2025wiv,
    author = "Santos, Felipe Bruno Medeiros dos and Morais, Jonathan and Pan, Supriya and Yang, Weiqiang and Di Valentino, Eleonora",
    title = "{A new window on dynamical dark energy: combining DESI-DR2 BAO with future gravitational wave observations}",
    eprint = "2504.04646",
    archivePrefix = "arXiv",
    primaryClass = "astro-ph.CO",
    doi = "10.1088/1475-7516/2026/07/022",
    journal = "JCAP",
    volume = "07",
    pages = "022",
    year = "2026"
}

@article{Specogna:2025guo,
    author = "Specogna, Enrico and Adil, Shahnawaz A. and Ozulker, Emre and Di Valentino, Eleonora and Nunes, Rafael C. and Akarsu, Ozgur and Sen, Anjan A.",
    title = "{Updated constraints on omnipotent dark energy: A comprehensive analysis with CMB and BAO data}",
    eprint = "2504.17859",
    archivePrefix = "arXiv",
    primaryClass = "gr-qc",
    doi = "10.1103/b7ht-lx26",
    journal = "Phys. Rev. D",
    volume = "113",
    number = "10",
    pages = "103549",
    year = "2026"
}

@article{Cheng:2025lod,
    author = "Cheng, Hanyu and Di Valentino, Eleonora and Escamilla, Luis A. and Sen, Anjan A. and Visinelli, Luca",
    title = "{Pressure parametrization of dark energy: first and second-order constraints with latest cosmological data}",
    eprint = "2505.02932",
    archivePrefix = "arXiv",
    primaryClass = "astro-ph.CO",
    reportNumber = "CA21106; CA21136",
    doi = "10.1088/1475-7516/2025/09/031",
    journal = "JCAP",
    volume = "09",
    pages = "031",
    year = "2025"
}

@article{Cheng:2025hug,
    author = "Cheng, Hanyu and Di Valentino, Eleonora and Visinelli, Luca",
    title = "{Cosmic strings as dynamical dark energy: Novel constraints}",
    eprint = "2505.22066",
    archivePrefix = "arXiv",
    primaryClass = "astro-ph.CO",
    doi = "10.1016/j.jheap.2026.100610",
    journal = "JHEAp",
    volume = "53",
    pages = "100610",
    year = "2026"
}

@article{Ozulker:2025ehg,
    author = {{\"O}z{\"u}lker, Emre and Di Valentino, Eleonora and Giar{\`e}, William},
    title = "{Dark Energy Crosses the Line: Quantifying and Testing the Evidence for Phantom Crossing}",
    eprint = "2506.19053",
    archivePrefix = "arXiv",
    primaryClass = "astro-ph.CO",
    month = "6",
    year = "2025"
}

@article{Li:2025vuh,
    author = "Li, Tian-Nuo and Du, Guo-Hong and Zhou, Sheng-Han and Li, Yun-He and Zhang, Jing-Fei and Zhang, Xin",
    title = "{Robust evidence for dynamical dark energy in light of DESI DR2 and joint ACT, SPT, and Planck data}",
    eprint = "2511.22512",
    archivePrefix = "arXiv",
    primaryClass = "astro-ph.CO",
    doi = "10.1016/j.dark.2026.102254",
    journal = "Phys. Dark Univ.",
    volume = "52",
    pages = "102254",
    year = "2026"
}

@article{Lee:2025pzo,
    author = "Lee, Dong Ha and Yang, Weiqiang and Di Valentino, Eleonora and Pan, Supriya and van de Bruck, Carsten",
    title = "{Shape of dark energy: Constraining its evolution with a general parametrization}",
    eprint = "2507.11432",
    archivePrefix = "arXiv",
    primaryClass = "astro-ph.CO",
    doi = "10.1103/z7y2-yvhg",
    journal = "Phys. Rev. D",
    volume = "113",
    number = "6",
    pages = "063554",
    year = "2026"
}

@article{Fazzari:2025lzd,
    author = "Fazzari, Elisa and Giar{\`e}, William and Di Valentino, Eleonora",
    title = "{Cosmographic Footprints of Dynamical Dark Energy}",
    eprint = "2509.16196",
    archivePrefix = "arXiv",
    primaryClass = "astro-ph.CO",
    doi = "10.3847/2041-8213/ae2917",
    journal = "Astrophys. J. Lett.",
    volume = "996",
    number = "1",
    pages = "L5",
    year = "2026"
}

@article{Smith:2025icl,
    author = {Smith, Adam and {\"O}z{\"u}lker, Emre and Di Valentino, Eleonora and van de Bruck, Carsten},
    title = "{Dynamical Dark Energy Meets Varying Electron Mass: Implications for Phantom Crossing and the Hubble Constant}",
    eprint = "2510.21931",
    archivePrefix = "arXiv",
    primaryClass = "astro-ph.CO",
    month = "10",
    year = "2025"
}

@article{Herold:2025hkb,
    author = "Herold, Laura and Karwal, Tanvi",
    title = "{Bayesian and frequentist perspectives agree on dynamical dark energy}",
    eprint = "2506.12004",
    archivePrefix = "arXiv",
    primaryClass = "astro-ph.CO",
    doi = "10.1103/gw6q-5k5j",
    journal = "Phys. Rev. D",
    volume = "113",
    number = "12",
    pages = "123551",
    year = "2026"
}

@article{Cheng:2025yue,
    author = "Cheng, Hanyu and Pan, Supriya and Di Valentino, Eleonora",
    title = "{Beyond Two Parameters: Revisiting Dark Energy with the Latest Cosmic Probes}",
    eprint = "2512.09866",
    archivePrefix = "arXiv",
    primaryClass = "astro-ph.CO",
    doi = "10.3847/1538-4357/ae3a8f",
    journal = "Astrophys. J.",
    volume = "999",
    number = "2",
    pages = "190",
    year = "2026"
}

@article{Gokcen:2026pkq,
    author = {G{\"o}k{\c{c}}en, Mine and Akarsu, {\"O}zg{\"u}r and Di Valentino, Eleonora},
    title = "{Revisiting CPL with sign-switching density: To cross or not to cross the NECB}",
    eprint = "2602.21169",
    archivePrefix = "arXiv",
    primaryClass = "astro-ph.CO",
    doi = "10.1016/j.dark.2026.102273",
    journal = "Phys. Dark Univ.",
    volume = "52",
    pages = "102273",
    year = "2026"
}

@article{Ishak:2025cay,
    author = "Ishak, Mustapha and Medina-Varela, Leonel",
    title = "{Persistent and serious challenge to the LCDM throne: Evidence for dynamical dark energy rising from combinations of different types of datasets}",
    eprint = "2507.22856",
    archivePrefix = "arXiv",
    primaryClass = "astro-ph.CO",
    month = "7",
    year = "2025"
}

@article{Najafi:2026kxs,
    author = "Najafi, Mahdi and Habibollahi, Mahdi and Reyhani, Masoume and Di Valentino, Eleonora and Pan, Supriya and Firouzjaee, Javad T. and Yang, Weiqiang",
    title = "{When Dark Energy Turns On: Constraints on a Critical Emergence Model}",
    eprint = "2603.13137",
    archivePrefix = "arXiv",
    primaryClass = "astro-ph.CO",
    month = "3",
    year = "2026"
}

@article{Yang:2026yaq,
    author = "Yang, Weiqiang and Di Valentino, Eleonora and Linder, Eric V. and Zhang, Sibo and Pan, Supriya",
    title = "{When One-Parameter Dark Energy Makes Neutrinos Physical Again}",
    eprint = "2603.15422",
    archivePrefix = "arXiv",
    primaryClass = "astro-ph.CO",
    month = "3",
    year = "2026"
}

@article{Kessler:2026dbi,
    author = "Kessler, Daniel A. and Di Valentino, Eleonora and Escamilla, Luis A. and Huterer, Dragan",
    title = "{Reconstructing dark energy with fewer assumptions}",
    eprint = "2606.05853",
    archivePrefix = "arXiv",
    primaryClass = "astro-ph.CO",
    month = "6",
    year = "2026"
}

@article{Lee:2026yzs,
    author = "Lee, Dong Ha and van de Bruck, Carsten and Di Valentino, Eleonora and Van Waerbeke, Ludovic and Zhitnitsky, Ariel",
    title = "{Evolving Dark Energy Is Vacuum Energy After All}",
    eprint = "2606.20036",
    archivePrefix = "arXiv",
    primaryClass = "astro-ph.CO",
    month = "6",
    year = "2026"
}

@article{Li:2026asg,
    author = "Li, Tian-Nuo and Du, Guo-Hong and Wang, Hao and Li, Yun-He and Zhang, Jing-Fei and Zhang, Xin",
    title = "{Dark Energy in the DESI Era: A Brief Review of Evidence, Beyond-{\ensuremath{\Lambda}}CDM Interpretations, and Tensions}",
    eprint = "2606.21826",
    archivePrefix = "arXiv",
    primaryClass = "astro-ph.CO",
    doi = "10.1088/1674-4527/ae8429",
    journal = "Res. Astron. Astrophys.",
    volume = "26",
    number = "8",
    pages = "084002",
    year = "2026"
}

@article{Giare:2026oti,
    author = "Giar{\`e}, William and Lee, Dong Ha and Di Valentino, Eleonora",
    title = "{Intertwined Constraints in Extended Cosmologies: Dark Energy, Curvature, Neutrinos, and Inflation}",
    eprint = "2607.01226",
    archivePrefix = "arXiv",
    primaryClass = "astro-ph.CO",
    month = "7",
    year = "2026"
}

@article{GuptaChoudhury:2026gsl,
    author = "Gupta Choudhury, Shibendu and Mukherjee, Purba and Di Valentino, Eleonora and Sen, Anjan A.",
    title = "{Model-Independent Indication for a Localized Anomaly in the Late-Time Expansion History}",
    eprint = "2607.13009",
    archivePrefix = "arXiv",
    primaryClass = "astro-ph.CO",
    month = "7",
    year = "2026"
}

@article{Amendola:1999er,
    author = "Amendola, Luca",
    title = "{Coupled quintessence}",
    eprint = "astro-ph/9908023",
    archivePrefix = "arXiv",
    doi = "10.1103/PhysRevD.62.043511",
    journal = "Phys. Rev. D",
    volume = "62",
    pages = "043511",
    year = "2000"
}

@article{Zimdahl:2001ar,
    author = "Zimdahl, Winfried and Pavon, Diego",
    title = "{Interacting quintessence}",
    eprint = "astro-ph/0105479",
    archivePrefix = "arXiv",
    doi = "10.1016/S0370-2693(01)01174-1",
    journal = "Phys. Lett. B",
    volume = "521",
    pages = "133--138",
    year = "2001"
}

@article{Farrar:2003uw,
    author = "Farrar, Glennys R. and Peebles, P. James E.",
    title = "{Interacting dark matter and dark energy}",
    eprint = "astro-ph/0307316",
    archivePrefix = "arXiv",
    doi = "10.1086/381728",
    journal = "Astrophys. J.",
    volume = "604",
    pages = "1--11",
    year = "2004"
}

@article{Wang:2016lxa,
    author = "Wang, B. and Abdalla, E. and Atrio-Barandela, F. and Pavon, D.",
    title = "{Dark Matter and Dark Energy Interactions: Theoretical Challenges, Cosmological Implications and Observational Signatures}",
    eprint = "1603.08299",
    archivePrefix = "arXiv",
    primaryClass = "astro-ph.CO",
    doi = "10.1088/0034-4885/79/9/096901",
    journal = "Rept. Prog. Phys.",
    volume = "79",
    number = "9",
    pages = "096901",
    year = "2016"
}

@article{Wang:2024vmw,
    author = "Wang, B. and Abdalla, E. and Atrio-Barandela, F. and Pav{\'o}n, D.",
    title = "{Further understanding the interaction between dark energy and dark matter: current status and future directions}",
    eprint = "2402.00819",
    archivePrefix = "arXiv",
    primaryClass = "astro-ph.CO",
    doi = "10.1088/1361-6633/ad2527",
    journal = "Rept. Prog. Phys.",
    volume = "87",
    number = "3",
    pages = "036901",
    year = "2024"
}

@article{DiValentino:2019ffd,
    author = "Di Valentino, Eleonora and Melchiorri, Alessandro and Mena, Olga and Vagnozzi, Sunny",
    title = "{Interacting dark energy in the early 2020s: A promising solution to the $H_0$ and cosmic shear tensions}",
    eprint = "1908.04281",
    archivePrefix = "arXiv",
    primaryClass = "astro-ph.CO",
    doi = "10.1016/j.dark.2020.100666",
    journal = "Phys. Dark Univ.",
    volume = "30",
    pages = "100666",
    year = "2020"
}

@article{Montani:2024pou,
    author = "Montani, Giovanni and Carlevaro, Nakia and Escamilla, Luis A. and Di Valentino, Eleonora",
    title = "{Kinetic model for dark energy{\textemdash}dark matter interaction: Scenario for the hubble tension}",
    eprint = "2404.15977",
    archivePrefix = "arXiv",
    primaryClass = "gr-qc",
    doi = "10.1016/j.dark.2025.101848",
    journal = "Phys. Dark Univ.",
    volume = "48",
    pages = "101848",
    year = "2025"
}

@article{vanderWesthuizen:2025rip,
    author = "van der Westhuizen, Marcel and Abebe, Amare and Di Valentino, Eleonora",
    title = "{III. Interacting Dark Energy: Summary of models, Pathologies, and Constraints}",
    eprint = "2509.04496",
    archivePrefix = "arXiv",
    primaryClass = "gr-qc",
    doi = "10.1016/j.dark.2025.102121",
    journal = "Phys. Dark Univ.",
    volume = "50",
    pages = "102121",
    year = "2025"
}

@article{Salvatelli:2013wra,
    author = "Salvatelli, Valentina and Marchini, Andrea and Lopez-Honorez, Laura and Mena, Olga",
    title = "{New constraints on Coupled Dark Energy from the Planck satellite experiment}",
    eprint = "1304.7119",
    archivePrefix = "arXiv",
    primaryClass = "astro-ph.CO",
    doi = "10.1103/PhysRevD.88.023531",
    journal = "Phys. Rev. D",
    volume = "88",
    number = "2",
    pages = "023531",
    year = "2013"
}

@article{Kumar:2016zpg,
    author = "Kumar, Suresh and Nunes, Rafael C.",
    title = "{Probing the interaction between dark matter and dark energy in the presence of massive neutrinos}",
    eprint = "1608.02454",
    archivePrefix = "arXiv",
    primaryClass = "astro-ph.CO",
    doi = "10.1103/PhysRevD.94.123511",
    journal = "Phys. Rev. D",
    volume = "94",
    number = "12",
    pages = "123511",
    year = "2016"
}

@article{Caprini:2016qxs,
    author = "Caprini, Chiara and Tamanini, Nicola",
    title = "{Constraining early and interacting dark energy with gravitational wave standard sirens: the potential of the eLISA mission}",
    eprint = "1607.08755",
    archivePrefix = "arXiv",
    primaryClass = "astro-ph.CO",
    doi = "10.1088/1475-7516/2016/10/006",
    journal = "JCAP",
    volume = "10",
    pages = "006",
    year = "2016"
}

@article{Murgia:2016ccp,
    author = "Murgia, Riccardo and Gariazzo, Stefano and Fornengo, Nicolao",
    title = "{Constraints on the Coupling between Dark Energy and Dark Matter from CMB data}",
    eprint = "1602.01765",
    archivePrefix = "arXiv",
    primaryClass = "astro-ph.CO",
    doi = "10.1088/1475-7516/2016/04/014",
    journal = "JCAP",
    volume = "04",
    pages = "014",
    year = "2016"
}

@article{Zheng:2017asg,
    author = "Zheng, Xiaogang and Biesiada, Marek and Cao, Shuo and Qi, Jingzhao and Zhu, Zong-Hong",
    title = "{Ultra-compact structure in radio quasars as a cosmological probe: a revised study of the interaction between cosmic dark sectors}",
    eprint = "1705.06204",
    archivePrefix = "arXiv",
    primaryClass = "astro-ph.CO",
    doi = "10.1088/1475-7516/2017/10/030",
    journal = "JCAP",
    volume = "10",
    pages = "030",
    year = "2017"
}

@article{Kumar:2017dnp,
    author = "Kumar, Suresh and Nunes, Rafael C.",
    title = "{Echo of interactions in the dark sector}",
    eprint = "1702.02143",
    archivePrefix = "arXiv",
    primaryClass = "astro-ph.CO",
    doi = "10.1103/PhysRevD.96.103511",
    journal = "Phys. Rev. D",
    volume = "96",
    number = "10",
    pages = "103511",
    year = "2017"
}

@article{DiValentino:2017iww,
    author = "Di Valentino, Eleonora and Melchiorri, Alessandro and Mena, Olga",
    title = "{Can interacting dark energy solve the $H_0$ tension?}",
    eprint = "1704.08342",
    archivePrefix = "arXiv",
    primaryClass = "astro-ph.CO",
    doi = "10.1103/PhysRevD.96.043503",
    journal = "Phys. Rev. D",
    volume = "96",
    number = "4",
    pages = "043503",
    year = "2017"
}

@article{Kumar:2021eev,
    author = "Kumar, Suresh",
    title = "{Remedy of some cosmological tensions via effective phantom-like behavior of interacting vacuum energy}",
    eprint = "2102.12902",
    archivePrefix = "arXiv",
    primaryClass = "astro-ph.CO",
    doi = "10.1016/j.dark.2021.100862",
    journal = "Phys. Dark Univ.",
    volume = "33",
    pages = "100862",
    year = "2021"
}

@article{Gao:2021xnk,
    author = "Gao, Li-Yang and Zhao, Ze-Wei and Xue, She-Sheng and Zhang, Xin",
    title = "{Relieving the H 0 tension with a new interacting dark energy model}",
    eprint = "2101.10714",
    archivePrefix = "arXiv",
    primaryClass = "astro-ph.CO",
    doi = "10.1088/1475-7516/2021/07/005",
    journal = "JCAP",
    volume = "07",
    pages = "005",
    year = "2021"
}

@article{Pan:2023mie,
    author = "Pan, Supriya and Yang, Weiqiang",
    title = "{On the interacting dark energy scenarios $-$ the case for Hubble constant tension}",
    eprint = "2310.07260",
    archivePrefix = "arXiv",
    primaryClass = "astro-ph.CO",
    doi = "10.1007/978-981-99-0177-7_29",
    month = "10",
    year = "2023"
}

@article{Benisty:2024lmj,
    author = "Benisty, David and Pan, Supriya and Staicova, Denitsa and Di Valentino, Eleonora and Nunes, Rafael C.",
    title = "{Late-time constraints on interacting dark energy: Analysis independent of H0, rd, and MB}",
    eprint = "2403.00056",
    archivePrefix = "arXiv",
    primaryClass = "astro-ph.CO",
    doi = "10.1051/0004-6361/202449883",
    journal = "Astron. Astrophys.",
    volume = "688",
    pages = "A156",
    year = "2024"
}

@article{Yang:2020uga,
    author = "Yang, Weiqiang and Di Valentino, Eleonora and Mena, Olga and Pan, Supriya and Nunes, Rafael C.",
    title = "{All-inclusive interacting dark sector cosmologies}",
    eprint = "2001.10852",
    archivePrefix = "arXiv",
    primaryClass = "astro-ph.CO",
    doi = "10.1103/PhysRevD.101.083509",
    journal = "Phys. Rev. D",
    volume = "101",
    number = "8",
    pages = "083509",
    year = "2020"
}

@article{Forconi:2023hsj,
    author = "Forconi, Matteo and Giar{\`e}, William and Mena, Olga and Ruchika and Di Valentino, Eleonora and Melchiorri, Alessandro and Nunes, Rafael C.",
    title = "{A double take on early and interacting dark energy from JWST}",
    eprint = "2312.11074",
    archivePrefix = "arXiv",
    primaryClass = "astro-ph.CO",
    doi = "10.1088/1475-7516/2024/05/097",
    journal = "JCAP",
    volume = "05",
    pages = "097",
    year = "2024"
}

@article{Pourtsidou:2016ico,
    author = "Pourtsidou, Alkistis and Tram, Thomas",
    title = "{Reconciling CMB and structure growth measurements with dark energy interactions}",
    eprint = "1604.04222",
    archivePrefix = "arXiv",
    primaryClass = "astro-ph.CO",
    doi = "10.1103/PhysRevD.94.043518",
    journal = "Phys. Rev. D",
    volume = "94",
    number = "4",
    pages = "043518",
    year = "2016"
}

@article{DiValentino:2020vnx,
    author = "Di Valentino, Eleonora",
    title = "{A combined analysis of the $H_0$ late time direct measurements and the impact on the Dark Energy sector}",
    eprint = "2011.00246",
    archivePrefix = "arXiv",
    primaryClass = "astro-ph.CO",
    reportNumber = "IPPP/20/72",
    doi = "10.1093/mnras/stab187",
    journal = "Mon. Not. Roy. Astron. Soc.",
    volume = "502",
    number = "2",
    pages = "2065--2073",
    year = "2021"
}

@article{DiValentino:2020leo,
    author = "Di Valentino, Eleonora and Mena, Olga",
    title = "{A fake Interacting Dark Energy detection?}",
    eprint = "2009.12620",
    archivePrefix = "arXiv",
    primaryClass = "astro-ph.CO",
    doi = "10.1093/mnrasl/slaa175",
    journal = "Mon. Not. Roy. Astron. Soc.",
    volume = "500",
    number = "1",
    pages = "L22--L26",
    year = "2020"
}

@article{Nunes:2021zzi,
    author = "Nunes, Rafael C. and Di Valentino, Eleonora",
    title = "{Dark sector interaction and the supernova absolute magnitude tension}",
    eprint = "2107.09151",
    archivePrefix = "arXiv",
    primaryClass = "astro-ph.CO",
    doi = "10.1103/PhysRevD.104.063529",
    journal = "Phys. Rev. D",
    volume = "104",
    number = "6",
    pages = "063529",
    year = "2021"
}

@article{Yang:2018uae,
    author = "Yang, Weiqiang and Mukherjee, Ankan and Di Valentino, Eleonora and Pan, Supriya",
    title = "{Interacting dark energy with time varying equation of state and the $H_0$ tension}",
    eprint = "1809.06883",
    archivePrefix = "arXiv",
    primaryClass = "astro-ph.CO",
    doi = "10.1103/PhysRevD.98.123527",
    journal = "Phys. Rev. D",
    volume = "98",
    number = "12",
    pages = "123527",
    year = "2018"
}

@article{Zhang:2018mlj,
    author = "Zhang, Jiajun and An, Rui and Luo, Wentao and Li, Zhaozhou and Liao, Shihong and Wang, Bin",
    title = "{The First Constraint from SDSS Galaxy{\textendash}Galaxy Weak Lensing Measurements on Interacting Dark Energy Models}",
    eprint = "1807.05522",
    archivePrefix = "arXiv",
    primaryClass = "astro-ph.CO",
    doi = "10.3847/2041-8213/ab133f",
    journal = "Astrophys. J. Lett.",
    volume = "875",
    number = "2",
    pages = "L11",
    year = "2019"
}

@article{vonMarttens:2019ixw,
    author = "von Marttens, Rodrigo and Lombriser, Lucas and Kunz, Martin and Marra, Valerio and Casarini, Luciano and Alcaniz, Jailson",
    title = "{Dark degeneracy I: Dynamical or interacting dark energy?}",
    eprint = "1911.02618",
    archivePrefix = "arXiv",
    primaryClass = "astro-ph.CO",
    doi = "10.1016/j.dark.2020.100490",
    journal = "Phys. Dark Univ.",
    volume = "28",
    pages = "100490",
    year = "2020"
}

@article{Lucca:2020zjb,
    author = "Lucca, Matteo and Hooper, Deanna C.",
    title = "{Shedding light on dark matter-dark energy interactions}",
    eprint = "2002.06127",
    archivePrefix = "arXiv",
    primaryClass = "astro-ph.CO",
    reportNumber = "ULB-TH/20-01",
    doi = "10.1103/PhysRevD.102.123502",
    journal = "Phys. Rev. D",
    volume = "102",
    number = "12",
    pages = "123502",
    year = "2020"
}

@article{Xiao:2021nmk,
    author = "Xiao, Linfeng and Costa, Andre A. and Wang, Bin",
    title = "{Forecasts on interacting dark energy from the 21-cm angular power spectrum with BINGO and SKA observations}",
    eprint = "2103.01796",
    archivePrefix = "arXiv",
    primaryClass = "astro-ph.CO",
    doi = "10.1093/mnras/stab3256",
    journal = "Mon. Not. Roy. Astron. Soc.",
    volume = "510",
    number = "1",
    pages = "1495--1514",
    year = "2021"
}

@article{Gao:2022ahg,
    author = "Gao, Li-Yang and Xue, She-Sheng and Zhang, Xin",
    title = "{Dark energy and matter interacting scenario to relieve H $_{0}$ and S $_{8}$ tensions*}",
    eprint = "2212.13146",
    archivePrefix = "arXiv",
    primaryClass = "astro-ph.CO",
    doi = "10.1088/1674-1137/ad2b52",
    journal = "Chin. Phys. C",
    volume = "48",
    number = "5",
    pages = "051001",
    year = "2024"
}

@article{Zhai:2023yny,
    author = "Zhai, Yuejia and Giar{\`e}, William and van de Bruck, Carsten and Di Valentino, Eleonora and Mena, Olga and Nunes, Rafael C.",
    title = "{A consistent view of interacting dark energy from multiple CMB probes}",
    eprint = "2303.08201",
    archivePrefix = "arXiv",
    primaryClass = "astro-ph.CO",
    doi = "10.1088/1475-7516/2023/07/032",
    journal = "JCAP",
    volume = "07",
    pages = "032",
    year = "2023"
}

@article{Joseph:2022khn,
    author = "Joseph, Albin and Saha, Rajib",
    title = "{Forecast analysis on interacting dark energy models from future generation PICO and DESI missions}",
    eprint = "2209.07167",
    archivePrefix = "arXiv",
    primaryClass = "astro-ph.CO",
    doi = "10.1093/mnras/stac3586",
    journal = "Mon. Not. Roy. Astron. Soc.",
    volume = "519",
    number = "2",
    pages = "1809--1822",
    year = "2022"
}

@article{Bernui:2023byc,
    author = "Bernui, Armando and Di Valentino, Eleonora and Giar{\`e}, William and Kumar, Suresh and Nunes, Rafael C.",
    title = "{Exploring the H0 tension and the evidence for dark sector interactions from 2D BAO measurements}",
    eprint = "2301.06097",
    archivePrefix = "arXiv",
    primaryClass = "astro-ph.CO",
    doi = "10.1103/PhysRevD.107.103531",
    journal = "Phys. Rev. D",
    volume = "107",
    number = "10",
    pages = "103531",
    year = "2023"
}

@article{Becker:2020hzj,
    author = {Becker, Niklas and Hooper, Deanna C. and Kahlhoefer, Felix and Lesgourgues, Julien and Sch{\"o}neberg, Nils},
    title = "{Cosmological constraints on multi-interacting dark matter}",
    eprint = "2010.04074",
    archivePrefix = "arXiv",
    primaryClass = "astro-ph.CO",
    reportNumber = "TTK-20-32, ULB-TH/20-13",
    doi = "10.1088/1475-7516/2021/02/019",
    journal = "JCAP",
    volume = "02",
    pages = "019",
    year = "2021"
}

@article{Hoerning:2023hks,
    author = "Hoerning, Gabriel A. and Landim, Ricardo G. and Ponte, Luiza O. and Rolim, Raphael P. and Abdalla, Filipe B. and Abdalla, Elcio",
    title = "{Constraints on interacting dark energy revisited: Implications for the Hubble tension}",
    eprint = "2308.05807",
    archivePrefix = "arXiv",
    primaryClass = "astro-ph.CO",
    doi = "10.1103/6zrh-8fmv",
    journal = "Phys. Rev. D",
    volume = "112",
    number = "2",
    pages = "023523",
    year = "2025"
}

@article{Giare:2024ytc,
    author = "Giar{\`e}, William and Zhai, Yuejia and Pan, Supriya and Di Valentino, Eleonora and Nunes, Rafael C. and van de Bruck, Carsten",
    title = "{Tightening the reins on nonminimal dark sector physics: Interacting dark energy with dynamical and nondynamical equation of state}",
    eprint = "2404.02110",
    archivePrefix = "arXiv",
    primaryClass = "astro-ph.CO",
    doi = "10.1103/PhysRevD.110.063527",
    journal = "Phys. Rev. D",
    volume = "110",
    number = "6",
    pages = "063527",
    year = "2024"
}

@article{Mukhopadhyay:2020bml,
    author = "Mukhopadhyay, Upala and Majumdar, Debasish and Datta, Kanan K.",
    title = "{Probing interacting dark energy and scattering of baryons with dark matter in light of the EDGES 21-cm signal}",
    eprint = "2008.09972",
    archivePrefix = "arXiv",
    primaryClass = "astro-ph.CO",
    doi = "10.1103/PhysRevD.103.063510",
    journal = "Phys. Rev. D",
    volume = "103",
    number = "6",
    pages = "063510",
    year = "2021"
}

@article{Escamilla:2023shf,
    author = "Escamilla, Luis A. and Akarsu, Ozgur and Di Valentino, Eleonora and Vazquez, J. Alberto",
    title = "{Model-independent reconstruction of the interacting dark energy kernel: Binned and Gaussian process}",
    eprint = "2305.16290",
    archivePrefix = "arXiv",
    primaryClass = "astro-ph.CO",
    doi = "10.1088/1475-7516/2023/11/051",
    journal = "JCAP",
    volume = "11",
    pages = "051",
    year = "2023"
}

@article{vanderWesthuizen:2023hcl,
    author = "van der Westhuizen, Marcel A. and Abebe, Amare",
    title = "{Interacting dark energy: clarifying the cosmological implications and viability conditions}",
    eprint = "2302.11949",
    archivePrefix = "arXiv",
    primaryClass = "gr-qc",
    doi = "10.1088/1475-7516/2024/01/048",
    journal = "JCAP",
    volume = "01",
    pages = "048",
    year = "2024"
}

@article{Silva:2024ift,
    author = "Silva, Emanuelly and Z{\'u}{\~n}iga-Bola{\~n}o, Ubaldo and Nunes, Rafael C. and Di Valentino, Eleonora",
    title = "{Non-linear matter power spectrum modeling in interacting dark energy cosmologies}",
    eprint = "2403.19590",
    archivePrefix = "arXiv",
    primaryClass = "astro-ph.CO",
    doi = "10.1140/epjc/s10052-024-13487-x",
    journal = "Eur. Phys. J. C",
    volume = "84",
    number = "10",
    pages = "1104",
    year = "2024"
}

@article{Zhao:2022ycr,
    author = "Zhao, Yu and Liu, Yun and Liao, Shihong and Zhang, Jiajun and Liu, Xiangkun and Du, Wei",
    title = "{Constraining interacting dark energy models with the halo concentration{\textendash}mass relation}",
    eprint = "2212.02050",
    archivePrefix = "arXiv",
    primaryClass = "astro-ph.CO",
    doi = "10.1093/mnras/stad1814",
    journal = "Mon. Not. Roy. Astron. Soc.",
    volume = "523",
    number = "4",
    pages = "5962--5971",
    year = "2023"
}

@article{Li:2024qso,
    author = "Li, Tian-Nuo and Wu, Peng-Ju and Du, Guo-Hong and Jin, Shang-Jie and Li, Hai-Li and Zhang, Jing-Fei and Zhang, Xin",
    title = "{Constraints on Interacting Dark Energy Models from the DESI Baryon Acoustic Oscillation and DES Supernovae Data}",
    eprint = "2407.14934",
    archivePrefix = "arXiv",
    primaryClass = "astro-ph.CO",
    doi = "10.3847/1538-4357/ad87f0",
    journal = "Astrophys. J.",
    volume = "976",
    number = "1",
    pages = "1",
    year = "2024"
}

@article{Pooya:2024wsq,
    author = "Pooya, N. Nazari",
    title = "{Growth of matter perturbations in the interacting dark energy-dark matter scenarios}",
    eprint = "2407.03766",
    archivePrefix = "arXiv",
    primaryClass = "astro-ph.CO",
    doi = "10.1103/PhysRevD.110.043510",
    journal = "Phys. Rev. D",
    volume = "110",
    number = "4",
    pages = "043510",
    year = "2024"
}

@misc{simplemc,
  author = "A. Slosar and J. A. Vazquez",
  year = 2020,
  howpublished = "\url{https://github.com/ja-vazquez/SimpleMC}"
}
\end{document}